\documentclass[12pt]{article}
\usepackage{a4wide,epsfig,epsf}

\newcommand{\agt}{\,\rlap{\lower 3.5 pt \hbox{$\mathchar \sim$}} \raise 1pt
 \hbox {$>$}\,}
\newcommand{\alt}{\,\rlap{\lower 3.5 pt \hbox{$\mathchar \sim$}} \raise 1pt
 \hbox {$<$}\,}

\catcode`@=11
\newcount\@tempcntc
\def\@citex[#1]#2{\if@filesw\immediate\write\@auxout{\string\citation{#2}}\fi
  \@tempcnta\z@\@tempcntb\m@ne\def\@citea{}\@cite{\@for\@citeb:=#2\do
    {\@ifundefined
       {b@\@citeb}{\@citeo\@tempcntb\m@ne\@citea\def\@citea{,}{\bf ?}\@warning
       {Citation `\@citeb' on page \thepage \space undefined}}%
    {\setbox\z@\hbox{\global\@tempcntc0\csname b@\@citeb\endcsname\relax}%
     \ifnum\@tempcntc=\z@ \@citeo\@tempcntb\m@ne
       \@citea\def\@citea{,}\hbox{\csname b@\@citeb\endcsname}%
     \else
      \advance\@tempcntb\@ne
      \ifnum\@tempcntb=\@tempcntc
      \else\advance\@tempcntb\m@ne\@citeo
      \@tempcnta\@tempcntc\@tempcntb\@tempcntc\fi\fi}}\@citeo}{#1}}
\def\@citeo{\ifnum\@tempcnta>\@tempcntb\else\@citea\def\@citea{,}%
  \ifnum\@tempcnta=\@tempcntb\the\@tempcnta\else
   {\advance\@tempcnta\@ne\ifnum\@tempcnta=\@tempcntb \else \def\@citea{--}\fi
    \advance\@tempcnta\m@ne\the\@tempcnta\@citea\the\@tempcntb}\fi\fi}
\catcode`@=12

\begin{document}

\title{
\vskip-3cm{\baselineskip14pt
\centerline{\normalsize DESY 04-149\hfill ISSN 0418-9833}
\centerline{\normalsize LPSC 04-040\hfill}
\centerline{\normalsize hep-ph/0408280\hfill}
\centerline{\normalsize August 2004\hfill}
}
\vskip1.5cm
$J/\psi$ plus prompt-photon associated production in two-photon collisions at
next-to-leading order}
\author{
{\sc M. Klasen,${}^{a,b}$ B.A. Kniehl,${}^b$ L.N. Mihaila,${}^b$
M. Steinhauser${}^b$}\\
{\normalsize ${}^a$ Universit\'e Grenoble I,
Laboratoire de Physique Subatomique et de Cosmologie,}\\
{\normalsize 53 Avenue des Martyrs, 38026 Grenoble, France}\\
{\normalsize ${}^b$ II. Institut f\"ur Theoretische Physik,
Universit\"at Hamburg,}\\
{\normalsize Luruper Chaussee 149, 22761 Hamburg, Germany}}

\date{}

\maketitle

\thispagestyle{empty}

\begin{abstract}
We calculate the cross section of $J/\psi$ plus prompt-photon inclusive
production in $\gamma\gamma$ collisions at next-to-leading order within the
factorization formalism of nonrelativistic quantum chromodynamics (NRQCD)
focusing on direct photoproduction.
Apart from direct $J/\psi$ production, we also include the feed-down from
directly-produced $\chi_{cJ}$ and $\psi^\prime$ mesons.
We discuss the analytical calculation, in particular the treatment of the
various types of singularities and the NRQCD operator renormalization, in some
detail.
We present theoretical predictions for the future $e^+e^-$ linear collider
TESLA, taking into account both brems- and beamstrahlung.

\medskip

\noindent
PACS numbers: 12.38.Bx, 12.39.St, 13.66.Bc, 14.40.Gx
\end{abstract}

\newpage

\section{Introduction}
\label{sec:one}

Since the discovery of the $J/\psi$ meson in 1974, charmonium has provided a
useful laboratory for quantitative tests of quantum chromodynamics (QCD) and,
in particular, of the interplay of perturbative and nonperturbative phenomena.
The factorization formalism of nonrelativistic QCD (NRQCD) \cite{cas,bbl}
provides a rigorous theoretical framework for the description of
heavy-quarkonium production and decay.
This formalism implies a separation of short-distance coefficients, which can 
be calculated perturbatively as expansions in the strong-coupling constant
$\alpha_s$, from long-distance matrix elements (MEs), which must be extracted
from experiment.
The relative importance of the latter can be estimated by means of velocity
scaling rules; {\it i.e.}, the MEs are predicted to scale with a definite
power of the heavy-quark ($Q$) velocity $v$ in the limit $v\ll1$.
In this way, the theoretical predictions are organized as double expansions in
$\alpha_s$ and $v$.
A crucial feature of this formalism is that it takes into account the complete
structure of the $Q\overline{Q}$ Fock space, which is spanned by the states
$n={}^{2S+1}L_J^{(a)}$ with definite spin $S$, orbital angular momentum
$L$, total angular momentum $J$, and color multiplicity $a=1,8$.
In particular, this formalism predicts the existence of color-octet (CO)
processes in nature.
This means that $Q\overline{Q}$ pairs are produced at short distances in
CO states and subsequently evolve into physical, color-singlet (CS) quarkonia
by the nonperturbative emission of soft gluons.
In the limit $v\to0$, the traditional CS model (CSM) \cite{ber} is recovered.
The greatest triumph of this formalism was that it was able to correctly 
describe \cite{bra} the cross section of inclusive charmonium
hadroproduction measured in $p\overline{p}$ collisions at the Fermilab
Tevatron \cite{abe}, which had turned out to be more than one order of
magnitude in excess of the theoretical prediction based on the CSM.
Apart from this phenomenological drawback, the CSM also suffers from severe
conceptual problems indicating that it is incomplete.
These include the presence of logarithmic infrared (IR) singularities in the
${\cal O}(\alpha_s)$ corrections to $P$-wave decays to light hadrons and in
the relativistic corrections to $S$-wave annihilation \cite{bar}, and the lack
of a general argument for its validity in higher orders of perturbation
theory.

In order to convincingly establish the phenomenological significance of the
CO processes, it is indispensable to identify them in other kinds of
high-energy experiments as well.
Studies of charmonium production in $ep$ photoproduction, $ep$ and $\nu N$
deep-inelastic scattering (DIS), $e^+e^-$ annihilation in the continuum,
$Z$-boson decays, $\gamma\gamma$ collisions, and $b$-hadron decays may be
found in the literature; for reviews, see Ref.~\cite{yua}.
Furthermore, the polarization of $\psi^\prime$ mesons produced directly and
of $J/\psi$ mesons produced promptly, {\it i.e.}, either directly or via the
feed-down from heavier charmonia, which also provides a sensitive probe of CO
processes, was investigated.
Until recently, none of these studies was able to prove or disprove the NRQCD
factorization hypothesis.
However, H1 data of $ep\to e+J/\psi+X$ in DIS at the DESY Hadron Electron Ring
Accelerator (HERA) \cite{h1} and DELPHI data of $\gamma\gamma\to J/\psi+X$ at
the CERN Large Electron Positron Collider (LEP2) \cite{delphi} provide first
independent evidence for it by agreeing with the respective NRQCD predictions
\cite{ep,gg}.

The verification of the NRQCD factorization hypothesis is presently hampered
both from the theoretical and experimental sides.
On the one hand, the theoretical predictions to be compared with existing
experimental data are, apart from very few exceptions \cite{kra,man,gre,kkms},
of lowest order (LO) and thus suffer from considerable uncertainties, mostly
from the dependences on the renormalization and factorization scales and from
the lack of information on the nonperturbative MEs.
On the other hand, the experimental errors are still rather sizeable.
The latter will be dramatically reduced with the upgrades of HERA (HERA II)
and the Tevatron (Run II) and with the advent of CERN LHC and hopefully a
future $e^+e^-$ linear collider (LC) such as the TeV-Energy Superconducting
Linear Accelerator (TESLA), which is presently being designed and planned at
DESY and has just been endorsed by the International Committee for Future
Accelerators (ICFA). 
On the theoretical side, it is necessary to calculate the
next-to-leading-order (NLO) corrections to the hard-scattering cross sections
and to include the effective operators which are suppressed by higher powers
in $v$.

In this paper, we concentrate on the inclusive production of $J/\psi$ mesons
in high-energy $\gamma\gamma$ collisions.
As mentioned above, this process was studied at LEP2 \cite{delphi}, where the
photons originated from hard initial-state bremsstrahlung.
At high-energy $e^+e^-$ LCs, an additional source of hard photons is provided
by beamstrahlung, the synchrotron radiation emitted by one of the colliding
bunches in the field of the opposite bunch.
The highest possible photon energies with large enough luminosity may be
achieved by converting the $e^+e^-$ LC into a $\gamma\gamma$ collider via
back-scattering of high-energetic laser light off the electron and positron
beams.

In order for a $J/\psi$ meson to acquire finite transverse momentum ($p_T$),
it must be produced together with another particle or a hadron jet ($j$).
Recently, we studied the process $\gamma\gamma\to J/\psi+j+X$, where $X$
denotes the hadronic remnant possibly including a second jet, at NLO
\cite{kkms}.
In this paper, we perform a similar analysis for the process
$\gamma\gamma\to J/\psi+\gamma+X$, where $\gamma$ represents a prompt photon.
This process leads to a spectacular signal.
On the one hand, $J/\psi$ mesons can be easily identified through their decays
to $e^+e^-$ or $\mu^+\mu^-$ pairs, with a combined branching fraction of
$B(J/\psi\to l^+l^-)=(11.81\pm0.14)\%$.
On the other hand, isolated energetic photons can be detected with high
efficiency through the distinctive showers they produce in the electromagnetic
calorimeter, which can be distinguished from those due to $\pi^0$ mesons on a
statistical basis, or through their conversion to $e^+e^-$ pairs.

The incoming photons can interact either directly with the quarks
participating in the hard-scattering process (direct photoproduction) or via
their quark and gluon content (resolved photoproduction).
Thus, the process $\gamma\gamma\to J/\psi+\gamma+X$ receives contributions
from the direct, single-resolved, and double-resolved channels.
All three contributions are formally of the same order in the perturbative
expansion.
This may be understood by observing that the parton density functions (PDFs)
of the photon have a leading behavior proportional to
$\alpha\ln(M^2/\Lambda_{\rm QCD}^2)\propto\alpha/\alpha_s$, where $\alpha$ is
the fine-structure constant, $M$ is the factorization scale, and
$\Lambda_{\rm QCD}$ is the asymptotic scale parameter of QCD.

In the following, we focus our attention on the direct channel, for which 
there is a phenomenological and a theoretical argument.
The former is exposed already at LO.
In fact, $\gamma\gamma\to J/\psi+\gamma+X$ almost exclusively proceeds through
direct photoproduction because it is then a CS process \cite{npb,jb}.
By contrast, the single-resolved contribution is strongly suppressed, by the
ratio
$\left\langle{\cal O}^{J/\psi}\left[{}^3\!S_1^{(8)}\right]\right\rangle/
\left\langle{\cal O}^{J/\psi}\left[{}^3\!S_1^{(1)}\right]\right\rangle\propto
v^4$.
There is also a CS process in the double-resolved channel.
However, it is mediated at the parton level via gluon fusion and thus 
suppressed by the smallness of the gluon PDF of the photon.
The theoretical argument is related to the observation that  
$\gamma\gamma\to J/\psi+\gamma+X$ via direct photoproduction does not yet
require mass factorization at NLO.
Therefore, this contribution does not suffer from an artificial $M$ 
dependence, so that it is justified to consider it separately.
By the same token, the $M$ dependence of the single-resolved contribution at 
LO is relatively feeble and to be compensated by the direct contribution at 
next-to-next-to-leading order (NNLO), which is beyond the scope of our
analysis.
The NLO treatments of the single-resolved and double-resolved channels are
left for future work.

In the parton model, prompt photons can also be produced through the
fragmentation of quarks and gluons, which necessitates the introduction of
photon fragmentation functions.
From arguments similar to those regarding the photon PDFs mentioned above, it
follows that also the photon fragmentation functions are formally of order
$\alpha/\alpha_s$.
Thus, consistency requires that their contribution be taken into account in
the theoretical description of prompt-photon inclusive production.
However, if the prompt-photon momentum is integrated over, then the
enhancement factor $\ln(M^2/\Lambda_{\rm QCD}^2)\propto1/\alpha_s$ disappears,
so that, in direct photoproduction, photon fragmentation is degraded to a
NNLO contribution.
Strictly speaking, we consider the process $\gamma\gamma\to J/\psi+X_\gamma$,
where $X_\gamma$ is a hadronic system containing a prompt photon, rather than
the process $\gamma\gamma\to J/\psi+\gamma+X$.

In order to distinguish prompt photons from bremsstrahlung photons, some
hardness condition must be introduced, {\it e.g.}, by requiring that the
transverse momentum of the photon be in excess of some minimum value.
From the theoretical point of view, this is essential because soft-photon
bremsstrahlung generates an IR catastrophe, which must be removed by including
also the corresponding virtual-photon contribution.
However, this would then be equivalent to calculating the ${\cal O}(\alpha)$
correction to the process $\gamma\gamma\to J/\psi+j$, which is suppressed
against the ${\cal O}(\alpha_s)$ one already provided in Ref.~\cite{kkms}, and
this is not our intention here.

The $J/\psi$ mesons can be produced directly; or via radiative or hadronic
decays of heavier charmonia, such as $\chi_{cJ}$ and $\psi^\prime$ mesons; or
via weak decays of $b$ hadrons.
The respective decay branching fractions are
$B(\chi_{c0}\to J/\psi+\gamma)=(1.18\pm0.14)\%$,
$B(\chi_{c1}\to J/\psi+\gamma)=(31.6\pm3.3)\%$,
$B(\chi_{c2}\to J/\psi+\gamma)=(20.2\pm1.7)\%$,
$B(\psi^\prime\to J/\psi+X)=(57.6\pm2.0)\%$, and
$B(B\to J/\psi+X)=(1.094\pm0.032)\%$ \cite{pdg}.
The $b$ hadrons can be detected by looking for displaced decay vertices with
dedicated vertex detectors, and the $J/\psi$ mesons originating from their
decays can thus be treated separately.
Therefore and because of the smallness of $B(B\to J/\psi+X)$, $J/\psi$
production through $b$-hadron decay is not considered here.
The cross sections of the four residual indirect production channels may be
approximated by multiplying the direct-production cross sections of the
respective intermediate charmonia with their decay branching fractions to
$J/\psi$ mesons.

To summarize, the goal of the present analysis is to calculate the inclusive
cross section of $\gamma\gamma\to J/\psi+X_\gamma$ in direct photoproduction
for finite values of $p_T$ at NLO within NRQCD allowing for the $J/\psi$ meson
to be promptly produced.
The LO result, also including the single- and double-resolved contributions,
may be found in Refs.~\cite{npb,jb} and the related references cited therein.
The leading relativistic correction, which originates from the $c\overline{c}$
Fock state $n={}^1\!P_1^{(1)}$ and is of ${\cal O}(v^4)$ relative to the LO
result, may be extracted from Ref.~\cite{npb}.
Apart from being of general phenomenological relevance,
the analyses reported here and in Ref.~\cite{kkms} should also be of
conceptual interest for the theoretical heavy-quarkonium community because
this is the first time that the full NLO corrections are evaluated for
inclusive $2\to2$ processes\footnote{%
Here, the $c\overline{c}$ bound state is considered as one particle.}
within the NRQCD framework.
Several theoretical problems that need to be tackled in our present analysis
were encountered in similar form in Ref.~\cite{kkms} and do not need to be
explained again in detail.
However, there are also qualitatively new features, especially in connection
with the operator renormalization and the cancellation of IR singularities,
which need to be discussed carefully.

This paper is organized as follows.
In Section~\ref{sec:ana}, we describe our analytical calculation in some
detail, focussing on the qualitatively new features.
In Section~\ref{sec:num}, we present our numerical results appropriate for the
$e^+e^-$ mode of TESLA, and discuss their phenomenological implications.
Our conclusions are summarized in Section~\ref{sec:con}.

\section{Analytic results}
\label{sec:ana}

We start this section with a few general remarks.
In our analytic calculation, we take the color gauge group to be SU($N_c$)
with a generic value of $N_c$, which is put equal to 3 in our numerical 
analysis.
Color factors appearing in our formulas include $T_F=1/2$,
$C_F=\left(N_c^2-1\right)/(2N_c)$, and $C_A=N_c$.
We work in the fixed-flavor-number scheme, with $n_f=3$ active quark flavors
$q=u,d,s$, which we treat as massless.
The charm quark $c$ and antiquark $\overline{c}$, with mass $m$, only appear
in the final state.
We denote the fractional electric charge of quark $q$ by $e_q$.

\renewcommand{\baselinestretch}{1.5}
\begin{table}[t]
\begin{center}
\begin{tabular}{|c|cc|}
\hline\hline
$k$ & $J/\psi$, $\psi^\prime$ & $\chi_{cJ}$ \\
\hline
3 & ${}^3\!S_1^{(1)}$ & --- \\
5 & --- & ${}^3\!P_J^{(1)}$, ${}^3\!S_1^{(8)}$ \\
7 & ${}^1\!S_0^{(8)}$, ${}^3\!S_1^{(8)}$, ${}^3\!P_J^{(8)}$ & --- \\
\hline\hline
\end{tabular}
\renewcommand{\baselinestretch}{1}
\caption{Values of $k$ in the velocity-scaling rule
$\left\langle{\cal O}^H[n]\right\rangle\propto v^k$ for the leading
$c\overline{c}$ Fock states $n$ pertinent to
$H=J/\psi,\chi_{cJ},\psi^\prime$.}
\label{tab:vsr}
\end{center}
\end{table}

The $c\overline{c}$ Fock states contributing at LO in $v$ are specified for
$H=J/\psi,\chi_{cJ},\psi^\prime$ in Table~\ref{tab:vsr}.
Their MEs satisfy the multiplicity relations
\begin{eqnarray}
\left\langle{\cal O}^{\psi(nS)}\left[{}^3\!P_J^{(8)}\right]\right\rangle
&=&(2J+1)
\left\langle{\cal O}^{\psi(nS)}\left[{}^3\!P_0^{(8)}\right]\right\rangle,
\nonumber\\
\left\langle{\cal O}^{\chi_{cJ}}\left[{}^3\!P_J^{(1)}\right]\right\rangle
&=&(2J+1)
\left\langle{\cal O}^{\chi_{c0}}\left[{}^3\!P_0^{(1)}\right]\right\rangle,
\nonumber\\
\left\langle{\cal O}^{\chi_{cJ}}\left[{}^3\!S_1^{(8)}\right]\right\rangle
&=&(2J+1)
\left\langle{\cal O}^{\chi_{c0}}\left[{}^3\!S_1^{(8)}\right]\right\rangle,
\label{eq:mul}
\end{eqnarray}
which follow to LO in $v$ from heavy-quark spin symmetry.

We employ dimensional regularization with $d=4-2\epsilon$ space-time
dimensions to handle the ultraviolet (UV) and IR singularities, and we
introduce unphysical 't~Hooft mass scales $\mu$ and $\lambda$ to ensure that
the renormalized strong-coupling constant and NRQCD MEs have the same mass
dimension as their LO counterparts.
We formally distinguish between UV and IR poles, which we denote as
$1/\epsilon_{\rm UV}$ and $1/\epsilon_{\rm IR}$, respectively.
We apply the projection method of Refs.~\cite{gre,kap}, which is equivalent to
the $d$-dimensional matching procedure of Ref.~\cite{bch}, in order to extract
the short-distance coefficients that multiply the MEs.
However, in order to conform with common standards, we adopt the
normalizations of the MEs from Ref.~\cite{bbl} rather than from
Refs.~\cite{man,gre}; {\it i.e.}, the MEs include spin and color average
factors.

There is only one partonic subprocess at LO, namely
\begin{equation}
\gamma(k_1)+\gamma(k_2)\to 
c\overline{c}\left[{}^3\!S_1^{(1)}\right](p)+\gamma(p^\gamma),
\label{eq:ccy}
\end{equation}
where the four-momentum assignments are indicated within the parentheses.
The corresponding Feynman diagrams are depicted in Fig.~\ref{fig:ccy}.
In the NLO analysis, we need to evaluate the cross section of process
(\ref{eq:ccy}) in $d$ space-time dimensions and retain terms of
${\cal O}(\epsilon)$ because UV counterterms appear in multiplicative
renormalization.
The final result is listed in Eqs.~(1) and (2) of Ref.~\cite{npb}.

The analogous process with $n={}^1\!P_1^{(1)}$ yields a relativistic 
correction of ${\cal O}(v^4)$, and its cross section may be obtained from
Eq.~(3) of Ref.~\cite{npb} by substituting
\begin{equation}
2N_cC_F\alpha_s\frac{
\left\langle{\cal O}^H\left[{}^1\!P_1^{(8)}\right]\right\rangle}
{N_c^2-1}\to
e_c^2\alpha\left\langle{\cal O}^H\left[{}^1\!P_1^{(1)}\right]\right\rangle.
\end{equation}

\subsection{Virtual corrections}

\begin{figure}[ht]
\begin{center}
\epsfxsize=\textwidth
\epsffile[10 350 420 430]{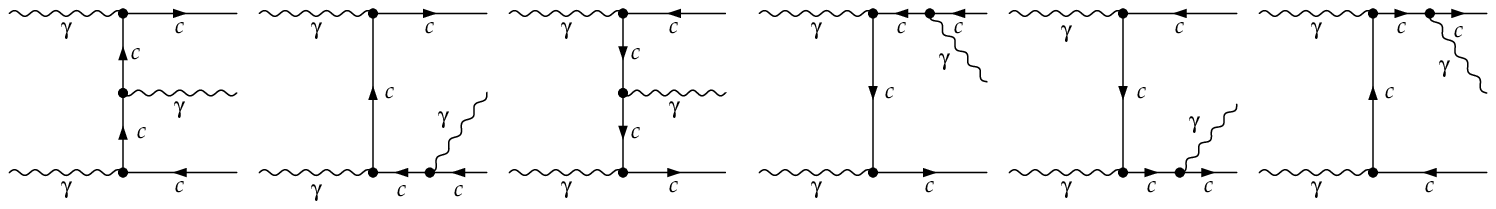}
\caption{Tree-level diagrams pertinent to the partonic
subprocess~(\ref{eq:ccy}).}
\label{fig:ccy}
\end{center}
\end{figure}

\begin{figure}[ht]
\begin{center}
\begin{tabular}{c}
\epsffile[10 350 420 500]{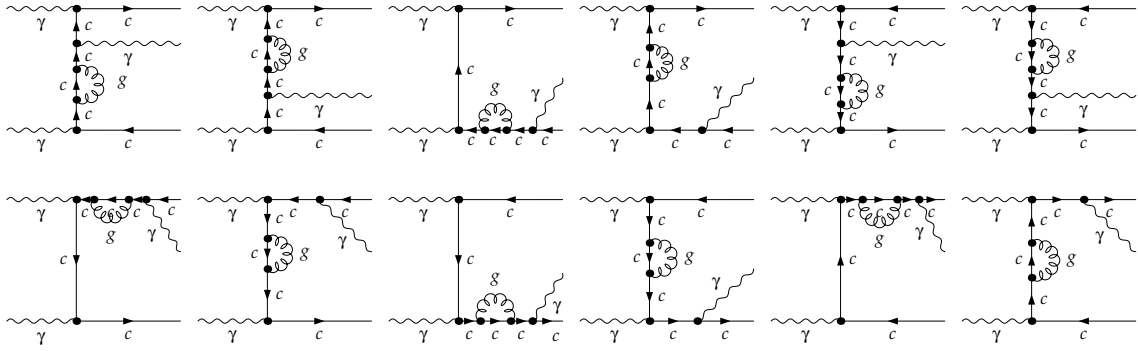}
\\
(a)
\end{tabular}
\caption{One-loop (a) self-energy, (b) triangle, (c) box, and (d) pentagon
diagrams pertinent to the partonic subprocess~(\ref{eq:ccy}).}
\label{fig:loop}
\end{center}
\end{figure}

The diagrams that generate the virtual corrections to the cross section of
process (\ref{eq:ccy}) are obtained by attaching one virtual gluon line in all
possible ways to the tree-level seed diagrams of Fig.~\ref{fig:ccy}.
They include the self-energy, triangle, box, and pentagon diagrams which are
depicted in Figs.~\ref{fig:loop}(a), (b), (c), and (d), respectively.
Loop insertions in external charm-quark lines are accommodated in the
corresponding wave-function renormalization constant and are not displayed in
Fig.~\ref{fig:loop}.
The self-energy and (non-vanishing) triangle diagrams are UV divergent; the
box diagrams are finite; and the pentagon diagrams are IR divergent.
The latter also contain Coulomb singularities, which are cancelled after
taking into account the corresponding corrections to the operator 
$\left\langle{\cal O}^H\left[{}^3\!S_1^{(1)}\right]\right\rangle$.
In the practical calculation, the Coulomb singularities are first regularized
by an infinitesimal gluon mass.
This regularization prescription is then transformed into one implemented with
a small relative velocity $v$ between the $c$ and $\overline{c}$ quarks
\cite{kra,Beenakker:2002nc}. 
In contrast to the case of $\gamma\gamma\to J/\psi+j+X$ considered in
Ref.~\cite{kkms}, diagrams containing a quark loop vanish by color
conservation.
Notice that, at NLO, virtual corrections only occur in connection with the
$c\overline{c}$ Fock state $n={}^3\!S_1^{(1)}$.

We apply two independent approaches to calculate the one-loop diagrams.
The first one uses {\tt FeynArts}~\cite{feynarts} to generate the diagrams and
self-written {\tt Mathematica} codes to apply the projectors and provide
expressions, which are afterwards treated with a {\tt FORM} program to perform
the tensor reduction and the extraction of the UV and IR singularities.
The result is then transformed into a {\tt Fortran} code to be used for the
numerical evaluation.
The second approach utilizes {\tt QGRAF}~\cite{Nogueira:1993ex} for the
generation of the diagrams, {\tt FeynCalc}~\cite{Mertig:an} for the tensor
reduction, and {\tt LoopTools}~\cite{looptools} for the numerical evaluation of
the IR-safe integrals.
These packages are surrounded by self-written interface programs, which allow
for a completely automated computation~\cite{Harlander:1998dq}.
Our analytic result is too lengthy to be presented here.

The self-energy and triangle diagrams of the type indicated in
Fig.~\ref{fig:loop} contain UV singularities, which are removed by
renormalizing the charm-quark mass $m$ and wave-function $\psi$ in the LO
diagrams of Fig.~\ref{fig:ccy}.
We renormalize $m$ and $\psi$ in the on-mass-shell scheme.
The relevant renormalization transformations may be found in Ref.~\cite{kkms}.

A crucial feature of effective field theories, such as NRQCD, is that the
composite operators are generally subject to renormalization.
In the case of NRQCD, this is essential in order to ensure the complete
cancellation of IR and Coulomb singularities at NLO and so to overcome the 
conceptual problems of the CSM mentioned in Section~\ref{sec:one}.
To be consistent with the rest of our calculation, we also employ dimensional
regularization here.
We adopt the technique described in Refs.~\cite{bbl,gre} to directly evaluate 
the NLO corrections to the NRQCD operators.
In this way, we avoid having to match partonic cross sections evaluated in
NRQCD with their counterparts in full QCD.

In the case under consideration, we have to renormalize the CS ME
$\left\langle{\cal O}^H\left[{}^3\!S_1^{(1)}\right]\right\rangle$, which
appears at LO.
In $d$ space-time dimensions, this ME has mass dimension $d-1$.
The 't~Hooft mass scale of NRQCD, $\lambda$, is introduced to keep its
renormalized version, which we wish to extract from experimental data, at mass
dimension 3.

The four-quark operator
$\left\langle{\cal O}^H\left[{}^3\!S_1^{(1)}\right]\right\rangle$ is related
to the amplitude for the elastic scattering of a $c\overline{c}$ pair. 
The corresponding tree-level diagram is depicted in Fig.~\ref{fig:op}(a).
The one-loop corrections to this amplitude are obtained by attaching a virtual
gluon line in all possible ways to the external heavy-quark legs, and they
involve self-energy and vertex corrections [see Fig.~\ref{fig:op}(b)--(e)].
Using the NRQCD Feynman rules in the quarkonium rest frame, expanding the
one-loop integrands as Taylor series in $1/m$, and performing the integration
over the loop momentum, we obtain the unrenormalized one-loop result
\begin{eqnarray}
\left\langle{\cal O}^H\left[{}^3\!S_1^{(1)}\right]\right\rangle_1
&=&\left\langle{\cal O}^H\left[{}^3\!S_1^{(1)}\right]\right\rangle_0
\left(1+\frac{\pi C_F\alpha_s}{2v}\right)
+\frac{2(C_A-2C_F)\alpha_s}{3\pi m^2}
\left(\frac{4\pi\mu^2}{\lambda^2}\right)^\epsilon\exp(-\epsilon\gamma_E)
\nonumber\\
&&{}\times\left(\frac{1}{\epsilon_{\rm UV}}-\frac{1}{\epsilon_{\rm IR}}\right)
\sum_{J=0}^2 \left\langle{\cal O}^H\left[{}^3\!P_J^{(8)}\right]\right\rangle,
\label{eq:reg}
\end{eqnarray}
where the subscript 0 labels the tree-level quantity, $\mu$ is the 't~Hooft
mass scale of QCD that enters through the $d$-dimensional loop integration,
and $\gamma_E$ is Euler's constant.
The term proportional to $1/v$ represents the Coulomb singularity, which
arises from the exchange of a longitudinal gluon between the outgoing $c$ and
$\overline{c}$ quarks [see Fig.~\ref{fig:op}(c)].
Obviously, NRQCD operators of different $c\overline{c}$ Fock states $n$ start
to mix at one loop.
Furthermore, the presence of UV singularities indicates that they need
renormalization.
In the following, we choose the $\overline{\rm MS}$ scheme for that.
We thus write
\begin{eqnarray}
\left\langle{\cal O}^H\left[{}^3\!S_1^{(1)}\right]\right\rangle_1
&=&\left\langle{\cal O}^H\left[{}^3\!S_1^{(1)}\right]\right\rangle_r(\lambda)
+\frac{2(C_A-2C_F)\alpha_s}{3\pi m^2}
\left(\frac{4\pi\mu^2}{\lambda^2}\right)^\epsilon\exp(-\epsilon\gamma_E)
\frac{1}{\epsilon_{\rm UV}}\nonumber\\
&&{}\times \sum_{J=0}^2
\left\langle{\cal O}^H\left[{}^3\!P_J^{(8)}\right]\right\rangle,
\label{eq:ct}
\end{eqnarray}
where the subscript $r$ labels the renormalized quantity and we identify
$\lambda$ with the NRQCD renormalization scale.
Inserting Eq.~(\ref{eq:ct}) into Eq.~(\ref{eq:reg}), we obtain
\begin{eqnarray}
\left\langle{\cal O}^H\left[{}^3\!S_1^{(1)}\right]\right\rangle_0
&=&\left\langle{\cal O}^H\left[{}^3\!S_1^{(1)}\right]\right\rangle_r(\lambda)
\left(1-\frac{\pi C_F\alpha_s}{2v}\right)
+\frac{2(C_A-2C_F)\alpha_s}{3\pi m^2}
\left(\frac{4\pi\mu^2}{\lambda^2}\right)^\epsilon\exp(-\epsilon\gamma_E)
\frac{1}{\epsilon_{\rm IR}}\nonumber\\
&&{}\times\sum_{J=0}^2
\left\langle{\cal O}^H\left[{}^3\!P_J^{(8)}\right]\right\rangle.
\label{eq:ren}
\end{eqnarray}
Substituting Eq.~(\ref{eq:ren}) into the LO result generates an IR counterterm
at ${\cal O}(\alpha_s)$ that is indispensible to render the total NLO result
finite.
The Coulomb singularity present in Eq.~(\ref{eq:ren}) is necessary to cancel
similar terms in the virtual corrections.

The renormalization group equation that determines the $\lambda$ dependence
of\break
$\left\langle{\cal O}^H\left[{}^3\!S_1^{(1)}\right]\right\rangle_r(\lambda)$
may be derived by differentiating Eq.~(\ref{eq:ren}) with respect to $\lambda$
and then taking the physical limit $\epsilon\to0$.
It reads
\begin{equation}
\frac{\lambda^2d}{d\lambda^2}
\left\langle{\cal O}^H\left[{}^3\!S_1^{(1)}\right]\right\rangle_r(\lambda)
=\frac{2(C_A-2C_F)\alpha_s(\mu)}{3\pi m^2}\sum_{J=0}^2
\left\langle{\cal O}^H\left[{}^3\!P_J^{(8)}\right]\right\rangle.
\label{eq:rge}
\end{equation}
There is no obvious physical reason to distinguish between the scales $\mu$
and $\lambda$, which refer to the $c\overline{c}g$ and
$c\overline{c}c\overline{c}$ vertices in the one-loop diagrams of
Figs.~\ref{fig:op}(b)--(e), respectively.
Both scales should essentially be of ${\cal O}(m)$.
In the following, we thus identify $\mu=\lambda$.
Integration of Eq.~(\ref{eq:rge}) then yields
\begin{equation}
\left\langle{\cal O}^H\left[{}^3\!S_1^{(1)}\right]\right\rangle_r(\lambda)
=\left\langle{\cal O}^H\left[{}^3\!S_1^{(1)}\right]\right\rangle_r(\lambda_0)
+\frac{8(C_A-2C_F)}{3m^2}\,
\frac{1}{\beta_0}\ln\frac{\alpha_s(\lambda_0)}{\alpha_s(\lambda)}
\sum_{J=0}^2\left\langle{\cal O}^H\left[{}^3\!P_J^{(8)}\right]\right\rangle,
\label{eq::oprun}
\end{equation}
where $\lambda_0$ is a reference scale for which the value of 
$\left\langle{\cal O}^H\left[{}^3\!S_1^{(1)}\right]\right\rangle_r(\lambda_0)$
is assumed to be known and $\beta_0=(11/3)C_A-(4/3)T_Fn_f$ is the one-loop
coefficient of the QCD beta function.
In want of a genuine NLO determination of the MEs
$\left\langle{\cal O}^H[n]\right\rangle_r(\lambda_0)$ from a fit to
experimental data, we choose $\lambda_0=m$ and identify
$\left\langle{\cal O}^H[n]\right\rangle_r(m)$ with their $\lambda$-independent
LO values, which are known from the literature \cite{bkl}.
Since the LO cross section is of ${\cal O}(\alpha_s^0)$, we have to employ in 
Eq.~(\ref{eq::oprun}) the one-loop formula for $\alpha_s(\lambda)$, which
reads
\begin{equation}
\alpha_s(\lambda)
=\frac{4\pi}{\beta_0\ln\left(\lambda^2/\Lambda_{\rm QCD}^2\right)}.
\label{eq:as}
\end{equation}

\subsection{Real corrections}

The real corrections to the cross section of process (\ref{eq:ccy}) arise from
the partonic subprocess
\begin{equation}
\gamma(k_1)+\gamma(k_2)\to c\overline{c}[n](p)+g(k_3)+\gamma(p^\gamma),
\label{eq:ccgy}
\end{equation}
where $n={}^1\!S_0^{(8)},{}^3\!P_J^{(8)}$.
The corresponding diagrams emerge by attaching one real gluon line in all
possible ways to the tree-level seed diagrams of Fig.~\ref{fig:ccy} and are
presented in Fig.~\ref{fig:ccgy}.
Process (\ref{eq:ccgy}) with $n={}^3\!S_1^{(1)},{}^3\!P_J^{(1)}$ is prohibited
by color conservation, since the $c\bar c$ pair must be in a CO state to
neutralize the color of the gluon.
On the other hand, process (\ref{eq:ccgy}) with $n={}^3\!S_1^{(8)}$ is
forbidden by Furry's theorem \cite{fur}, as may be understood by observing
that the ${}^3\!S_1^{(8)}$ projector effectively closes the charm-quark line
and acts like a vector coupling, so that we are dealing with a closed fermion
loop containing five vector couplings.
This was also verified by explicit calculation.

Generically denoting the transition matrix ($T$) elements of process
(\ref{eq:ccgy}) by ${\cal T}_r$, its cross sections may be evaluated as
\begin{equation}
d\sigma_r=\frac{1}{2s}d{\rm PS}_3(k_1+k_2;p,k_3,p^\gamma)
\overline{|{\cal T}_r|^2},
\label{eq:xsr}
\end{equation}
where $s=(k_1+k_2)^2$ is the square of the center-of-mass energy.
Here and in the following, we denote the Lorentz-invariant $N$-particle
phase-space element in $d$ dimensions as
\begin{equation}
d{\rm PS}_N(P;p_1,\ldots,p_N)=\mu^{(N-1)(4-d)}
(2\pi)^d\delta^{(d)}\left(P-\sum_{i=1}^Np_i\right)
\prod_{i=1}^N\frac{d^dp_i}{(2\pi)^{d-1}}\delta\left(p_i^2-m_i^2\right)
\theta\left(p_i^0\right),
\end{equation}
where $p_i$ and $m_i$ are the four-momenta and masses of the outgoing
particles and $P$ is the total four-momentum of the incoming particles.

Integrating Eq.~(\ref{eq:xsr}) over the three-particle phase space while
keeping the value of $p_T$ finite and constraining the final-state photon to
be hard, we encounter IR singularities of the soft type in the case of
$n={}^3\!P_J^{(8)}$.
In fact, in the soft-gluon limit, $k_3\to0$, $\overline{|{\cal T}_r|^2}$
factorizes as
\begin{equation}
\overline{|{\cal T}_r|^2}=\overline{|{\cal T}_0|^2}F_3,
\end{equation}
where ${\cal T}_0$ is the $T$-matrix element of process (\ref{eq:ccy}) and
\begin{equation}
F_3=\frac{8\pi(C_A-2C_F)\alpha_s}{3(p\!\cdot\!k_3)^2}
\left(1-\frac{\epsilon}{3}+{\cal O}(\epsilon^2)\right)
\end{equation}
is the appropriate Eikonal factor.
Since the soft gluon does not affect the kinematics of the hard process
involving the residual particles, the three-particle phase space can be
decomposed as
\begin{equation}
d{\rm PS}_3(k_1+k_2;p,k_3,p^\gamma)=d{\rm PS}_2(k_1+k_2;p,p^\gamma)
d{\rm PS}_s^3,
\end{equation}
where
\begin{equation}
d{\rm PS}_s^3=\mu^{4-d}\frac{d^{d-1}k_3}{(2\pi)^{d-1}2k_3^0}.
\end{equation}
The integration of the Eikonal factor over the soft part of the phase space,
can be performed analytically and yields \cite{harris}
\begin{eqnarray}
I_s^3&=&\int_{k_3^0<\delta\sqrt s/2}d{\rm PS}_s^3F_3
\nonumber\\
&=&-\frac{(C_A-2C_F)\alpha_s}{3\pi m^2}
\left(\frac{4\pi\mu^2}{\delta^2s}\right)^\epsilon\exp(-\epsilon\gamma_E)
\left(\frac{1}{\epsilon}
+\frac{1}{\beta}\ln\frac{1+\beta}{1-\beta}-\frac{1}{3}
+{\cal O}(\epsilon)\right),
\end{eqnarray}
where $\delta$ is an infinitesimal dimensionless phase-space-slicing parameter
and\break $\beta=\left(s-4m^2\right)/\left(s+4m^2\right)$.

Since, in the diagrams of Fig.~\ref{fig:ccgy}, all photons and gluons are
coupled to the massive $c$-quark line, IR singularities of the collinear type
do not occur.
The integration of Eq.~(\ref{eq:xsr}) over the complementary region of phase
space can be performed numerically in $d=4$ space-time dimensions because
there are no UV or IR singularities.

\subsection{Assembly of the NLO cross section}

The NLO result for the cross section of process~(\ref{eq:ccy}) is obtained by
adding the virtual and real corrections to the LO result.
Collecting the various contributions discussed above, arising from the virtual
corrections (vi), the parameter and wave-function renormalization (ct), the
operator redefinition (op), the soft-gluon radiation (so), and the hard-gluon
emission (ha), we can schematically write the resulting differential cross
section of $\gamma\gamma\to H+X_\gamma$, including the MEs, as
\begin{eqnarray}
d\sigma(\mu,\lambda)&=&
d\sigma_0(\lambda)[1
+\delta_{\rm vi}(\mu;\epsilon_{\rm UV},\epsilon_{\rm IR},v)
+\delta_{\rm ct}(\mu;\epsilon_{\rm UV},\epsilon_{\rm IR})
+\delta_{\rm op}(\mu,\lambda;\epsilon_{\rm IR},v)]\nonumber\\
&&{}+d\sigma_{\rm so}(\mu,\lambda;\epsilon_{\rm IR},\delta)
+d\sigma_{\rm ha}(\mu,\lambda;\delta),
\label{eq:sum}
\end{eqnarray}
where the dependences on the unphysical mass scales $\mu$ and $\lambda$, and
the regulators $\epsilon_{\rm IR}$, $\epsilon_{\rm UV}$, $v$, and $\delta$ are
indicated in parentheses for each term.
Everywhere in Eq.~(\ref{eq:sum}), $\alpha_s(\mu)$ is evaluated in the
$\overline{\rm MS}$ scheme using the one-loop formula~(\ref{eq:as}), $m$ is
defined in the OS scheme, and the MEs are understood as their
$\overline{\rm MS}$ values $\langle{\cal O}^H[n]\rangle_r(\lambda)$.
This is necessary to ensure the exact cancellation of the singularities.

The right-hand side of Eq.~(\ref{eq:sum}) is manifestly finite.
The UV divergences cancel between $\delta_{\rm vi}$ and $\delta_{\rm ct}$;
the IR singularities among $\delta_{\rm vi}$, $\delta_{\rm ct}$,
$\delta_{\rm op}$, and $d\sigma_{\rm so}$; and the Coulomb singularities
between $\delta_{\rm vi}$ and $\delta_{\rm op}$.
Note that $\delta_{\rm op}$ is UV finite upon operator renormalization;
therefore, $\delta_{\rm op}$ in Eq.~(\ref{eq:sum}) does not depend any more on
$\epsilon_{\rm UV}$.
The full details about the cancellation of the IR and Coulomb singularities in
the various $c\overline{c}$ Fock states $n$ are presented in
Table~\ref{tab:ir}.

\renewcommand{\baselinestretch}{1.5}
\begin{table}[ht]
\begin{center}
\begin{tabular}{|l|l|l|}
\hline\hline
Subprocess & Source & IR- or Coulomb-singular term \\
\hline
$\gamma+\gamma\to c\bar{c}\left[{}^3\!S_1^{(1)}\right]+\gamma$
& virtual &
$\frac{\pi C_F\alpha_s}{2v}
\left(\frac{4\pi\mu^2}{m^2}\right)^\epsilon\exp(-\epsilon\gamma_E)
\overline{|{\cal T}_0|^2}$
\\
& operator &
$-\frac{\pi C_F\alpha_s}{2v}
\left\langle{\cal O}^H\left[{}^3\!S_1^{(1)}\right]\right\rangle$
\\
$\gamma+\gamma\to c\overline{c}\left[{}^3\!P_J^{(8)}\right]+g+\gamma$
& operator &
$\frac{2(C_A-2C_F)\alpha_s}{3\pi m^2}\,\frac{1}{\epsilon}
\left(\frac{4\pi\mu^2}{\lambda^2}\right)^\epsilon\exp(-\epsilon\gamma_E)
\left\langle{\cal O}^H\left[{}^3\!P_J^{(8)}\right]\right\rangle$
\\
& soft &
$-\frac{2(C_A-2C_F)\alpha_s}{3\pi m^2}\,\frac{1}{\epsilon}
\left(\frac{4\pi\mu^2}{m^2}\right)^\epsilon\exp(-\epsilon\gamma_E)
\overline{|{\cal T}_0|^2}$
\\
\hline\hline
\end{tabular}
\renewcommand{\baselinestretch}{1}
\caption{Compilation of the IR- and Coulomb-singular terms arising from the
various sources in the various partonic subprocesses.}
\label{tab:ir}
\end{center}
\end{table}

The right-hand side of Eq.~(\ref{eq:sum}) is independent of the cut-off
parameter $\delta$.
The $\delta$ dependence cancels between $d\sigma_{\rm so}$ and
$d\sigma_{\rm ha}$.
While $d\sigma_{\rm so}$ is known in analytic form, the phase-space integrals
occurring in the evaluation of $d\sigma_{\rm ha}$ are rather cumbersome and
are thus solved numerically.
Consequently, the cancellation of $\delta$ has to be established numerically.
The goodness of this cancellation is assessed in Section~\ref{sec:num}.

The right-hand side of Eq.~(\ref{eq:sum}) is independent of the
renormalization scales $\mu$ and $\lambda$ up to terms that are formally
beyond NLO;
this cancellation is not exact because the running of $\alpha_s(\mu)$ and
$\langle{\cal O}^H[n]\rangle_r(\lambda)$ is determined from the respective
renormalization group equations, which resum logarithmic corrections to all
orders.

\section{Numerical results}
\label{sec:num}

We are now in a position to present our numerical analysis of the inclusive
production of prompt $J/\psi$ mesons in association with prompt photons in
two-photon collisions with direct photon interactions at NLO in the NRQCD
factorization framework.
We consider TESLA in its $e^+e^-$ mode with $\sqrt s=500$~GeV, where the
photons are produced via bremsstrahlung and beamstrahlung.

We first specify our input parameters.
We use $m=1.5$~GeV, $\alpha=1/137.036$ \cite{pdg}, and the one-loop formula for
$\alpha_s^{(n_f)}(\mu)$ given in Eq.~(\ref{eq:as}), with $n_f=3$ active quark
flavors and $\Lambda_{\rm QCD}^{(3)}=204$~MeV \cite{grs,mrst}.
Our default choice of renormalization scales is $\mu=m_T$, where
$m_T=\sqrt{p_T^2+4m^2}$ is the transverse mass of the $J/\psi$ meson, and
$\lambda=m$.
For the LO evaluation of the single- and double-resolved contributions, we use
the LO set of photon PDFs from Gl\"uck, Reya, and Schienbein (GRS) \cite{grs},
which are the only available ones that are implemented in the
fixed-flavor-number scheme, with $n_f=3$, and identify the factorization
scale with $\mu$.
In want of NLO sets of $J/\psi$, $\chi_{cJ}$, and $\psi^\prime$ MEs, we adopt
the LO sets determined in Ref.~\cite{bkl} using the LO set of proton PDFs from
Martin, Roberts, Stirling, and Thorne (MRST98LO) \cite{mrst}.
Specifically,
$\left\langle{\cal O}^{\psi(nS)}\left[{}^3\!S_1^{(1)}\right]\right\rangle$ and
$\left\langle{\cal O}^{\chi_{c0}}\left[{}^3\!P_0^{(1)}\right]\right\rangle$
were extracted from the measured partial decay widths of $\psi(nS)\to l^+l^-$
and $\chi_{c2}\to\gamma\gamma$ \cite{pdg}, respectively, while
$\left\langle{\cal O}^{\psi(nS)}\left[{}^1\!S_0^{(8)}\right]\right\rangle$,
$\left\langle{\cal O}^{\psi(nS)}\left[{}^3\!S_1^{(8)}\right]\right\rangle$,
$\left\langle{\cal O}^{\psi(nS)}\left[{}^3\!P_0^{(8)}\right]\right\rangle$,
and
$\left\langle{\cal O}^{\chi_{c0}}\left[{}^3\!S_1^{(8)}\right]\right\rangle$
were fitted to the transverse-momentum distributions of $\psi(nS)$ and
$\chi_{cJ}$ inclusive hadroproduction \cite{abe} and the cross-section ratio
$\sigma_{\chi_{c2}}/\sigma_{\chi_{c1}}$ \cite{aff} measured at the Tevatron.
The fit results for
$\left\langle{\cal O}^{\psi(nS)}\left[{}^1\!S_0^{(8)}\right]\right\rangle$ and
$\left\langle{\cal O}^{\psi(nS)}\left[{}^3\!P_0^{(8)}\right]\right\rangle$ are
strongly correlated, so that the linear combination
\begin{equation}
M_r^{\psi(nS)}
=\left\langle{\cal O}^{\psi(nS)}\left[{}^1\!S_0^{(8)}\right]\right\rangle
+\frac{r}{m^2}
\left\langle{\cal O}^{\psi(nS)}\left[{}^3\!P_0^{(8)}\right]\right\rangle,
\label{eq:mr}
\end{equation}
with a suitable value of $r$, is quoted.
Unfortunately, Eq.~(\ref{eq:sum}) is sensitive to linear combination of
$\left\langle{\cal O}^{\psi(nS)}\left[{}^1\!S_0^{(8)}\right]\right\rangle$ and
$\left\langle{\cal O}^{\psi(nS)}\left[{}^3\!P_0^{(8)}\right]\right\rangle$
that is different from the one appearing in Eq.~(\ref{eq:mr}).
In want of more specific information, we thus make the democratic choice
$\left\langle{\cal O}^{\psi(nS)}\left[{}^1\!S_0^{(8)}\right]\right\rangle
=\left(r/m^2\right)
\left\langle{\cal O}^{\psi(nS)}\left[{}^3\!P_0^{(8)}\right]\right\rangle
=M_r^{\psi(nS)}/2$.

We now discuss the photon flux functions that enter our predictions for
photoproduction in the $e^+e^-$ mode of TESLA.
The energy spectrum of the bremsstrahlung photons is well described in the
Weizs\"acker-Williams approximation (WWA) \cite{wwa} by Eq.~(27) of
Ref.~\cite{fri}.
We assume that the scattered electrons and positrons will be antitagged, as
was usually the case at LEP2, and take the maximum scattering angle to be
$\theta_{\rm max}=25$~mrad \cite{theta}.
The energy spectrum of the beamstrahlung photons is approximately described by
Eq.~(2.14) of Ref.~\cite{pes}.
It is controlled by the effective beamstrahlung parameter $\Upsilon$, which is
given by Eq.~(2.10) of that reference.
Inserting the relevant TESLA parameters for the $\sqrt S=500$~GeV baseline
design specified in Table~1.3.1 of Ref.~\cite{tesla1} in that formula, we
obtain $\Upsilon=0.053$.
We coherently superimpose the WWA and beamstrahlung spectra.

At TESLA, prompt photons with scattering angles $7^\circ<\theta<173^\circ$ are
expected to be detectable, as can be read off from Fig.~2.4.2(a) in
Ref.~\cite{tesla2}.
This corresponds to the pseudorapidity window $|y^\gamma|<2.79$.
As a typical discrimination criterion against bremsstrahlung photons, we
enforce the hardness condition $p_T^\gamma>p_{T,\,{\rm min}}^\gamma$ with
$p_{T,\,{\rm min}}^\gamma=2m=3$~GeV.
This choice can be justified by studying the $p_{T,\,{\rm min}}^\gamma$
variation (see Fig.~\ref{fig:pty} below).
Unless otherwise stated, we always impose these acceptance cuts on
$p_T^\gamma$ and $y^\gamma$ in the following.

With the NLO corrections to single- and double-resolved photoproduction yet to
be evaluated, we are not in a position to present a complete phenomenological
prediction that could be confronted with experimental data as it stands.
Therefore, we refrain from presenting a full-fledged quantitative estimate of
the theoretical uncertainties.
However, a first indication of their size may be obtained by investigating the
dependences of our numerical evaluation on the renormalization scales $\mu$
and $\lambda$.
For this purpose, it is sufficient to consider a typical kinematic situation.
We thus choose as our reference quantity the differential cross section
$d^2\sigma/dp_T\,dy$ at $p_T=5$~GeV and $y=0$.

In Fig.~\ref{fig:mu}, the NLO result (solid line) is shown as a function of
$\mu$ for $m_T/4<\mu<4m_T$, while $\lambda$ is kept fixed at its reference
value.
For comparison, the LO result (dashed line), which is of course $\mu$
independent, is also shown. 
As expected, we observe that the $\mu$ dependence of the NLO result is rather
sizeable, reflecting the unscreened appearance of $\alpha_s(\mu)$ as an
overall factor of the radiative correction.
A partial compensation of this $\mu$ dependence can only occur at
next-to-next-to-leading order, which is, however, beyond the scope of this
work.
On the other hand, it is commonly believed that the size of unknown
higher-order corrections can be estimated by scale variations over judiciously
chosen ranges of values.
In this sense, we read off from Fig.~\ref{fig:mu} that the NLO result varies
by ${+20\atop-32}\%$ over the interval $m_T/2<\mu<2m_T$.
We also observe that the QCD correction is negative for all values of $\mu$
considered here.
This feature is scrutinized below, in connection with Figs.~\ref{fig:xs} and
\ref{fig:k}.

In Fig.~\ref{fig:la}, the NLO result (solid line) is shown as a function of
$\lambda$ for $m/2<\lambda<4m$, while $\mu$ is kept fixed at its reference
value.
Since
$\left\langle{\cal O}^H\left[{}^1\!S_0^{(8)}\right]\right\rangle_r(\lambda)$
and
$\left\langle{\cal O}^H\left[{}^3\!P_0^{(8)}\right]\right\rangle_r(\lambda)$
only enter Eq.~(\ref{eq:sum}) at NLO, through $d\sigma_{\rm so}$ and
$d\sigma_{\rm ha}$, we keep them at $\lambda=m$.
Then, the $\lambda$-dependent terms in Eq.~(\ref{eq:sum}) are just $d\sigma_0$
and $\delta_{\rm op}$.
In order to exhibit the partial compensation in $\lambda$ dependence between
these two terms, we also include in Fig.~\ref{fig:la} the results that are
obtained by only varying $\lambda$ in $d\sigma_0$ (dashed line) or
$\delta_{\rm op}$ (dotted line) at a time.
The dotted line is straight, reflecting the fact that $\delta_{\rm op}$
depends on $\lambda$ through a single logarithm.
The theoretical uncertainty related to the $\lambda$ variation in the interval
$m/2<\lambda<2m$ amounts to ${+2.1\atop-0.4}\%$.

In the remainder of this section, we explore the phenomenological consequences
of our analysis by studying the size and impact of the NLO corrections on the
$p_T^\gamma>p_{T,\,{\rm min}}^\gamma$, $p_T$ and $y$ distributions of the
cross section.
In Fig.~\ref{fig:pty}, we examine $d^2\sigma/dp_T\,dy$ for $p_T=5$~GeV, $y=0$,
and $|y^\gamma|<2.79$ as a function of $p_{T,\,{\rm min}}^\gamma$ at LO
(dashed line) and NLO (solid line).
At LO, the prompt photon balances the transverse momentum of the $J/\psi$
meson, so that we integrate over a delta function located at
$p_T^\gamma=p_T$ as long as $p_{T,\,{\rm min}}^\gamma<p_T$, which explains the
step-function shape of the $p_{T,\,{\rm min}}^\gamma$ distribution.
On the other hand, at NLO, there are events with an additional gluon-initiated
hadron jet ($j$) to share the recoil transverse momentum of the $J/\psi$ with
the prompt photon, so that, in particular, the latter attains access to the
phase-space region where $p_T^\gamma>p_T$.
Of course, such events are suppressed by the smallness of $\alpha_s$ and, for
$p_T^\gamma\gg p_T$, also by phase-space limitation.
This explains the size and shape of the tail in the NLO result for
$p_T^\gamma>p_T$.

In Fig.~\ref{fig:xs}, we study $d^2\sigma/dp_T\,dy$ (a) for $y=0$ as a
function of $p_T$ and (b) for $p_T=5$~GeV as a function of $y$.
In each case, the LO (dashed line) and NLO (solid line) results of direct
photoproduction are shown.
In Fig.~\ref{fig:xs}(a), the LO results of single-resolved (dotted line) and
double-resolved (dot-dashed line) photoproduction are displayed as well.
Notice that our analysis is only valid for finite values of $p_T$; in the
limit $p_T\to0$, additional IR singularities occur, which require a more
sophisticated scheme of phase space slicing.
Therefore, we do not consider $p_T$ values below 2~GeV in 
Fig.~\ref{fig:xs}(a).

As explained in the context of Fig.~\ref{fig:pty}, we have $p_T=p_T^\gamma$ at
LO, which explains the sharp threshold at $p_T=p_{T,\,{\rm min}}^\gamma$ in
the dashed, dotted, and dot-dashed curves in Fig.~\ref{fig:xs}(a).
By contrast, the corresponding threshold in the solid curve in
Fig.~\ref{fig:xs}(a) is washed out by $J/\psi\gamma j$ events.
Comparing the LO results in Fig.~\ref{fig:xs}(a), we observe that the
direct-photoproduction contribution is overwhelming.
It exceeds the single-resolved one, which is exclusively generated by the CO
partonic subprocess
$\gamma+g\to c\overline{c}\left[{}^3\!S_1^{(8)}\right]+\gamma$, by more than
three orders of magnitude, essentially reflecting the ME ratio
$\left\langle{\cal O}^{J/\psi}\left[{}^3\!S_1^{(8)}\right]\right\rangle
/\left\langle{\cal O}^{J/\psi}\left[{}^3\!S_1^{(1)}\right]\right\rangle
\approx 3\times10^{-3}$.
For $p_T\alt6$~GeV, the double-resolved contribution is chiefly produced by
the CS partonic subprocess
$g+g\to c\overline{c}\left[{}^3\!S_1^{(1)}\right]+\gamma$, but it is
nevertheless substantially suppressed by two factors of the gluon PDF of the
photon and thus significantly undershoots the direct one, by approximately
three orders of magnitude.
At large values of $p_T$, the CO partonic subprocess
$q+\overline{q}\to c\overline{c}\left[{}^3\!S_1^{(8)}\right]+\gamma$ takes
over the lead.
This is typical for {\it fragmentation-prone} partonic subprocesses \cite{jb},
which contain a gluon with small virtuality, $q^2=4m^2$, that splits into a
$c\overline{c}$ pair in the Fock state $n={}^3\!S_1^{(8)}$ and thus generally
generate dominant contributions at $p_T\gg2m$ due to the presence of a large
gluon propagator.
The relative importance of the three photoproduction modes is likely to be
subject to change, at least at large values of $p_T$, once the as-yet unknown
NLO corrections for the single- and double-resolved contributions will be
included.
In particular, the NLO correction to the single-resolved contribution is
expected to be sizeable because fragmentation-prone partonic subprocesses,
which are absent at LO, start to contribute.
In fact, a similar situation was encountered in Ref.~\cite{kkms}.

From Figs.~\ref{fig:xs}(a) and (b), we observe that the NLO result of direct
photoproduction undershoots the LO one wherever the latter is non-zero,
except at some edges of phase space.
The magnitude of this reduction increases with the value of $p_T$ and
decreases towards the forward and backward directions.
This is nicely illustrated in Figs.~\ref{fig:k}, where the NLO to LO ratio
$K$ (solid line) is shown (a) for $y=0$ as a function of $p_T$ and (b) for
$p_T=5$~GeV as a function of $y$, respectively.

At first sight, the steady fall-off of the $K$ factor as a function of $p_T$
in Fig.~\ref{fig:k}(a) is surprising, and it is interesting to identify its
origin.
To this end, we break up the $K$ factor into the contributions related to the
$c\overline{c}$ Fock states
$n={}^3\!S_1^{(1)},{}^1\!S_0^{(8)},{}^3\!P_J^{(8)}$ as
\begin{equation}
K=1+\delta\left[{}^3\!S_1^{(1)}\right]
+\delta\left[{}^1\!S_0^{(8)}\right]
+\delta\left[{}^3\!P_J^{(8)}\right],
\label{eq:k}
\end{equation}
where $\delta\left[{}^3\!S_1^{(1)}\right]$ is due to the virtual correction,
and $\delta\left[{}^1\!S_0^{(8)}\right]$ and
$\delta\left[{}^3\!P_J^{(8)}\right]$ stem from the real correction.
Since the parts of the NLO correction that enter
$\delta\left[{}^3\!S_1^{(1)}\right]$,
$\delta\left[{}^1\!S_0^{(8)}\right]$, and
$\delta\left[{}^3\!P_J^{(8)}\right]$  stem from different $c\overline{c}$ Fock
states, they are separately gauge independent.
However, the first and third of them are IR divergent, so that we are led to
consider their $\overline{\rm MS}$-subtracted finite remainders.
In order not to artificially introduce large logarithms, we choose the
subtraction points to be $\mu=m_T$ and $\lambda=m$, respectively.
In order to illustrate how the $K$ factor in Eq.~(\ref{eq:k}) is gradually
built up, the results for $1+\delta\left[{}^3\!S_1^{(1)}\right]$ and
$1+\delta\left[{}^3\!S_1^{(1)}\right]+\delta\left[{}^1\!S_0^{(8)}\right]$ are
also shown in Figs.~\ref{fig:k}(a) and (b), as the dashed and dotted curves,
respectively.
We observe from Figs.~\ref{fig:k}(a) and (b) that the $K$ factor is dominated
by $\delta\left[{}^3\!S_1^{(1)}\right]$, while
$\delta\left[{}^1\!S_0^{(8)}\right]$ and $\delta\left[{}^3\!P_J^{(8)}\right]$
are of minor importance, which reflects the proportion of the respective MEs,
$\left\langle{\cal O}^{J/\psi}\left[{}^3\!S_1^{(1)}\right]\right\rangle$,
$\left\langle{\cal O}^{J/\psi}\left[{}^1\!S_0^{(8)}\right]\right\rangle$,
$\left\langle{\cal O}^{J/\psi}\left[{}^3\!P_J^{(8)}\right]\right\rangle$,
appearing as overall factors.
In fact, we have
$\left\langle{\cal O}^{J/\psi}\left[{}^3\!S_1^{(1)}\right]\right\rangle/
M_{3.4}^{J/\psi}\approx15$ \cite{bkl}.
We also notice that $\delta\left[{}^1\!S_0^{(8)}\right]$ is always positive,
while the $\overline{\rm MS}$-subtracted quantity
$\delta\left[{}^3\!P_J^{(8)}\right]$ is negative for $p_T\alt8$~GeV.
We conclude that $\delta\left[{}^3\!S_1^{(1)}\right]$ is responsible for the
striking fall-off of the $K$ factor with increasing values of $p_T$.

\section{Conclusions}
\label{sec:con}

We calculated the inclusive cross section of
$\gamma\gamma\to J/\psi+X_\gamma$, where the system $X_\gamma$ contains a
prompt photon, in direct photoproduction for finite values of $p_T$ at NLO
within NRQCD allowing for the $J/\psi$ meson to be promptly produced, and
presented phenomenological predictions for TESLA in the $e^+e^-$ mode of
operation.
At LO, $\gamma\gamma\to J/\psi+X_\gamma$ proceeds almost exclusively though
direct photoproduction because this is a CS process.
Its cross section is sizeable and its signal spectacular.
These observations provided a solid motivation for our work.

As for the real corrections, we employed the phase-space slicing method to
demarcate the regions of phase space containing soft singularities from the
hard regions, where the phase-space integrations were carried out numerically.
We verified that the combined result is, to very good approximation,
independent of the choice of the cut-off parameter $\delta$, over an extended
range of values.
We worked in dimensional regularization in connection with the
$\overline{\rm MS}$ renormalization and factorization schemes, so that our NLO
result depends on the QCD and NRQCD renormalization scales $\mu$ and
$\lambda$, respectively.
While the $\lambda$ dependence is formally cancelled up to terms beyond NLO,
the $\mu$ dependence is unscreened and introduces an appreciable theoretical
uncertainty.
Our NLO result does not involve any factorization scale and is thus formally
independent of single-resolved photoproduction and photon fragmentation.

We found that the inclusion of the NLO correction leads to a substantial
reduction in cross section, especially towards large values of $p_T$ and in
the central region of the $y$ range.
In fact, the $K$ factor falls below 0.1 for $p_T\agt12$~GeV.
These features could be traced to the ($\overline{\rm MS}$-subtracted) virtual
correction, which makes up the bulk of the NLO correction.

In order to complete the NLO treatment of $\gamma\gamma\to J/\psi+X_\gamma$,
we still need to evaluate the NLO corrections to single- and double-resolved
photoproduction.
Due to the occurrence of fragmentation-prone partonic subprocesses, these are
expected to lead to a substantial enhancement in the case of single-resolved
photoproduction.
Nevertheless, it is unlikely that the dominance of direct photoproduction will
be challenged at NLO.
Then, also $J/\psi+X_\gamma$ production in photoproduction at HERA and
hadroproduction at the Tevatron can be described at NLO.
This will help to provide a solid basis for an ultimate test of the NRQCD
factorization framework.

\bigskip
\noindent
{\bf Acknowledgements}
\smallskip

M.K. thanks the $2^{\rm nd}$ Institute for Theoretical Physics at the
University of Hamburg for the hospitality extended to him during a visit when
this manuscript was finalized.
This work was supported in part by the Deutsche Forschungsgemeinschaft through
Grant No.\ KN~365/1-2, by the Bundesministerium f\"ur Bildung und Forschung
through Grant No.\ 05~HT4GUA/4, and by Sun Microsystems through Academic
Equipment Grant No.~EDUD-7832-000332-GER.

\newpage
 
\newpage

\newpage
\begin{figure}[ht]
\begin{center}
\begin{tabular}{c}
\epsffile[10 70 420 430]{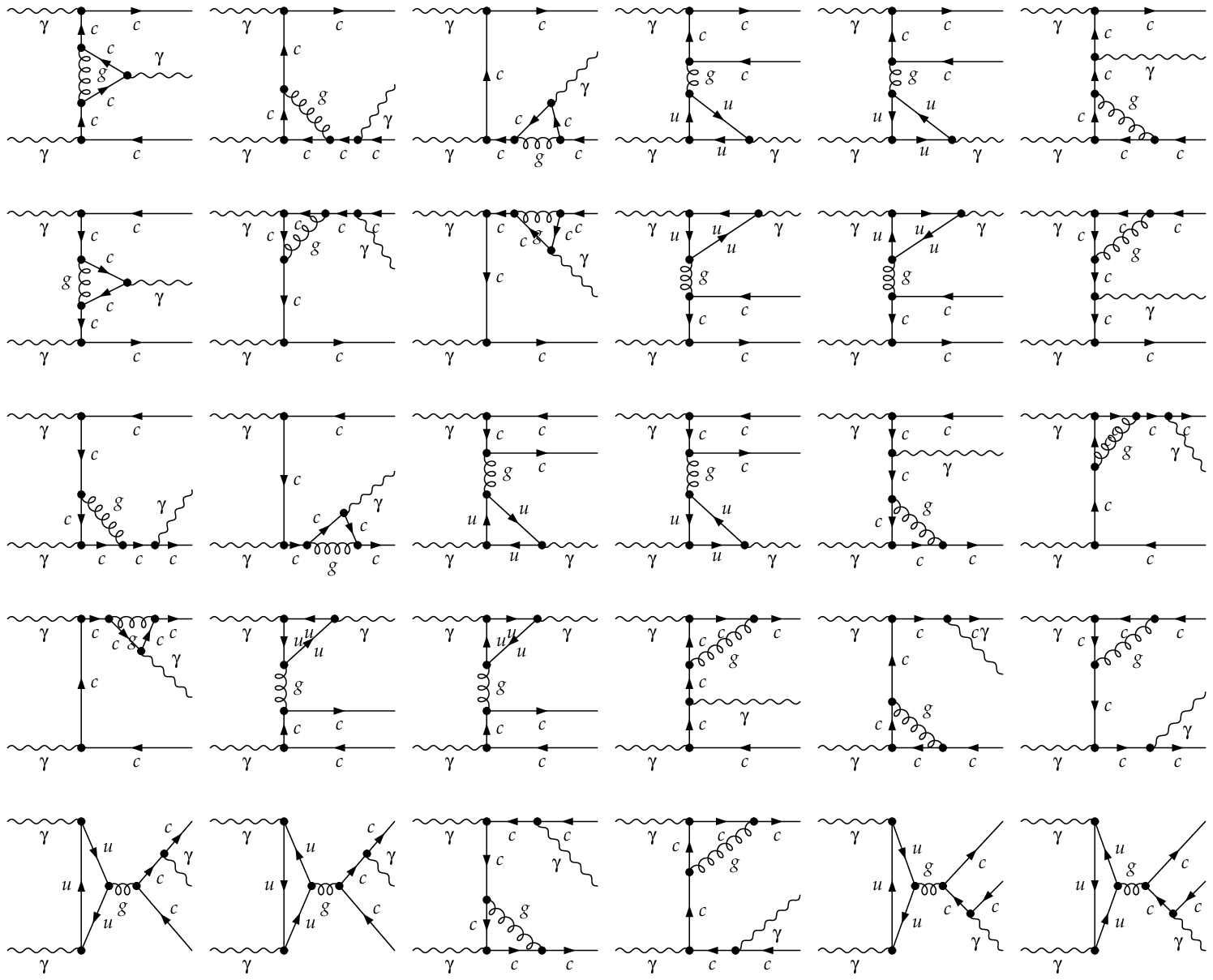}
\\
(b)
\end{tabular}
\\
Fig.~\ref{fig:loop} (continued).
\end{center}
\end{figure}

\newpage
\begin{figure}[ht]
\begin{center}
\begin{tabular}{c}
\epsffile[10 200 420 430]{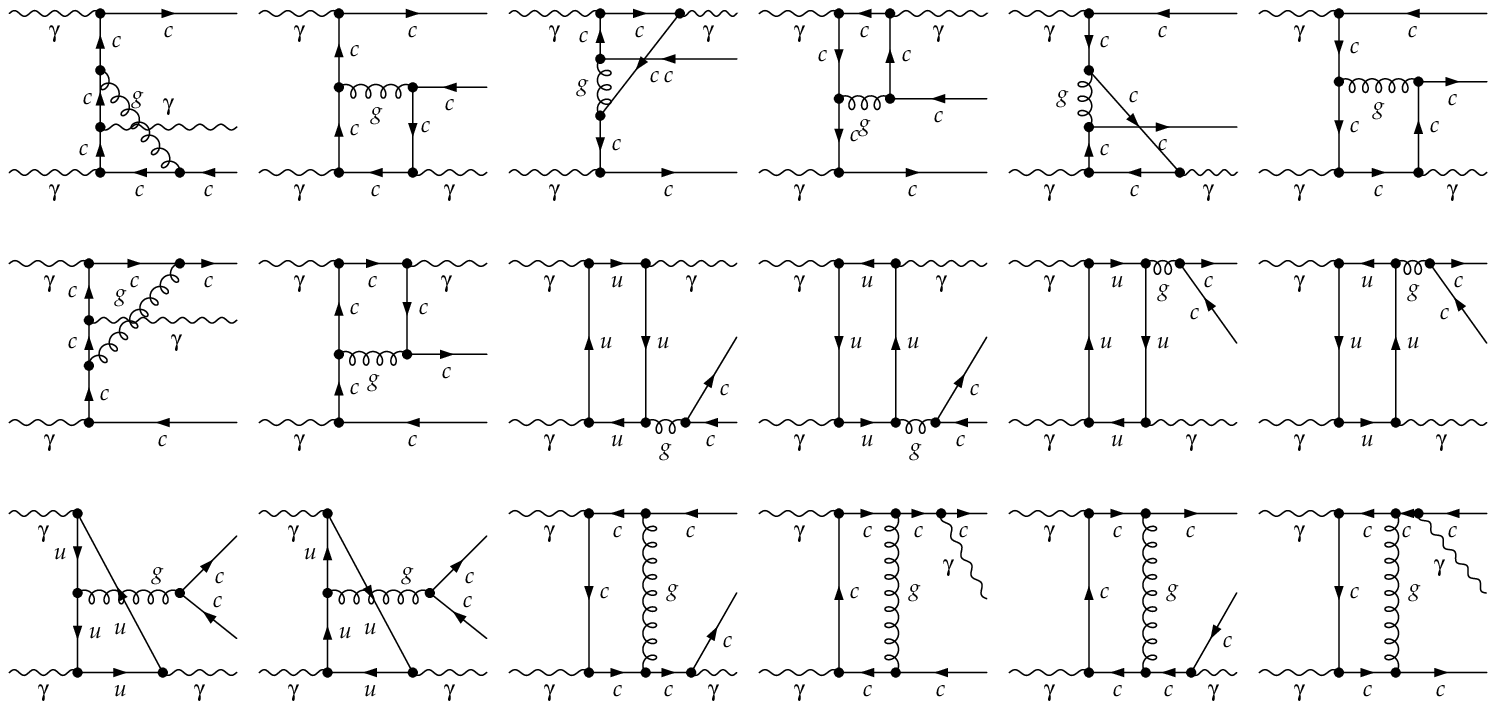}
\\
(c)
\\
\epsffile[10 350 420 430]{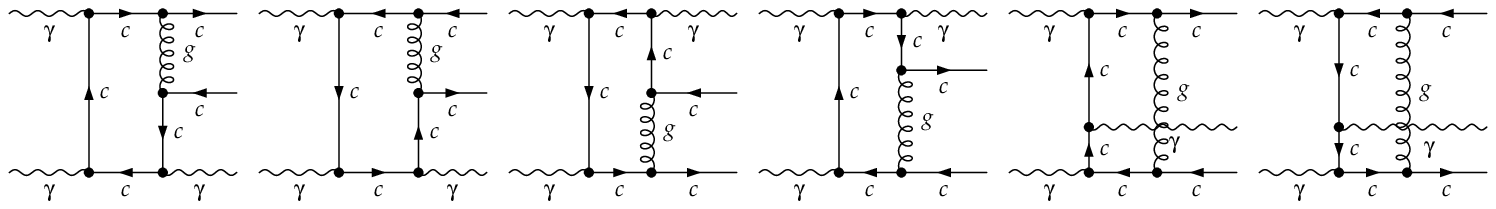}
\\
(d)
\end{tabular}
\\
Fig.~\ref{fig:loop} (continued).
\end{center}
\end{figure}

\newpage
\begin{figure}[ht]
\begin{center}
\epsfig{figure=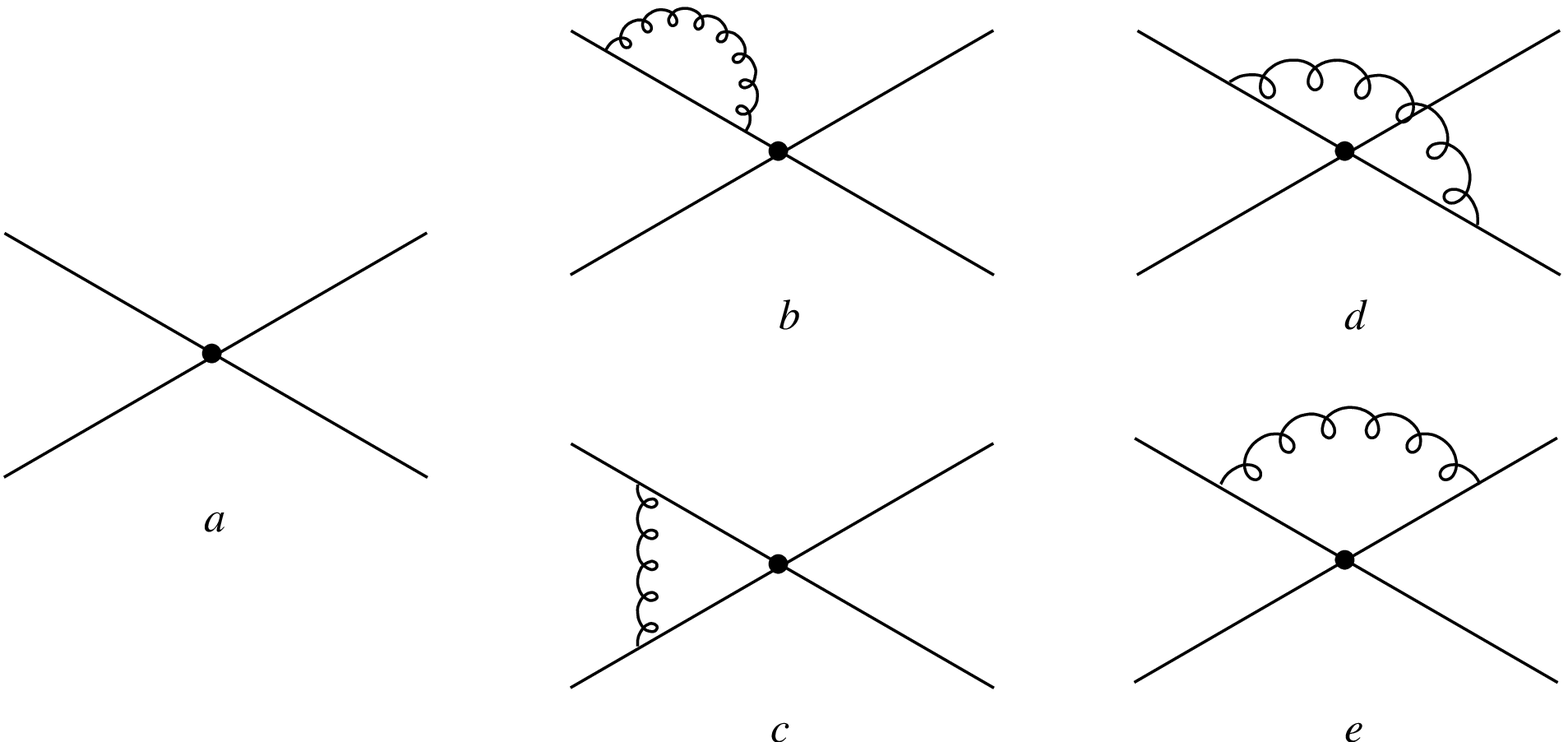,width=\textwidth}
\caption{Diagrams pertinent to the one-loop correction to
$\left\langle{\cal O}^H\left[{}^3\!S_1^{(1)}\right]\right\rangle$.}
\label{fig:op}
\end{center}
\end{figure}

\newpage
\begin{figure}[ht]
\begin{center}
\epsffile[10 150 420 430]{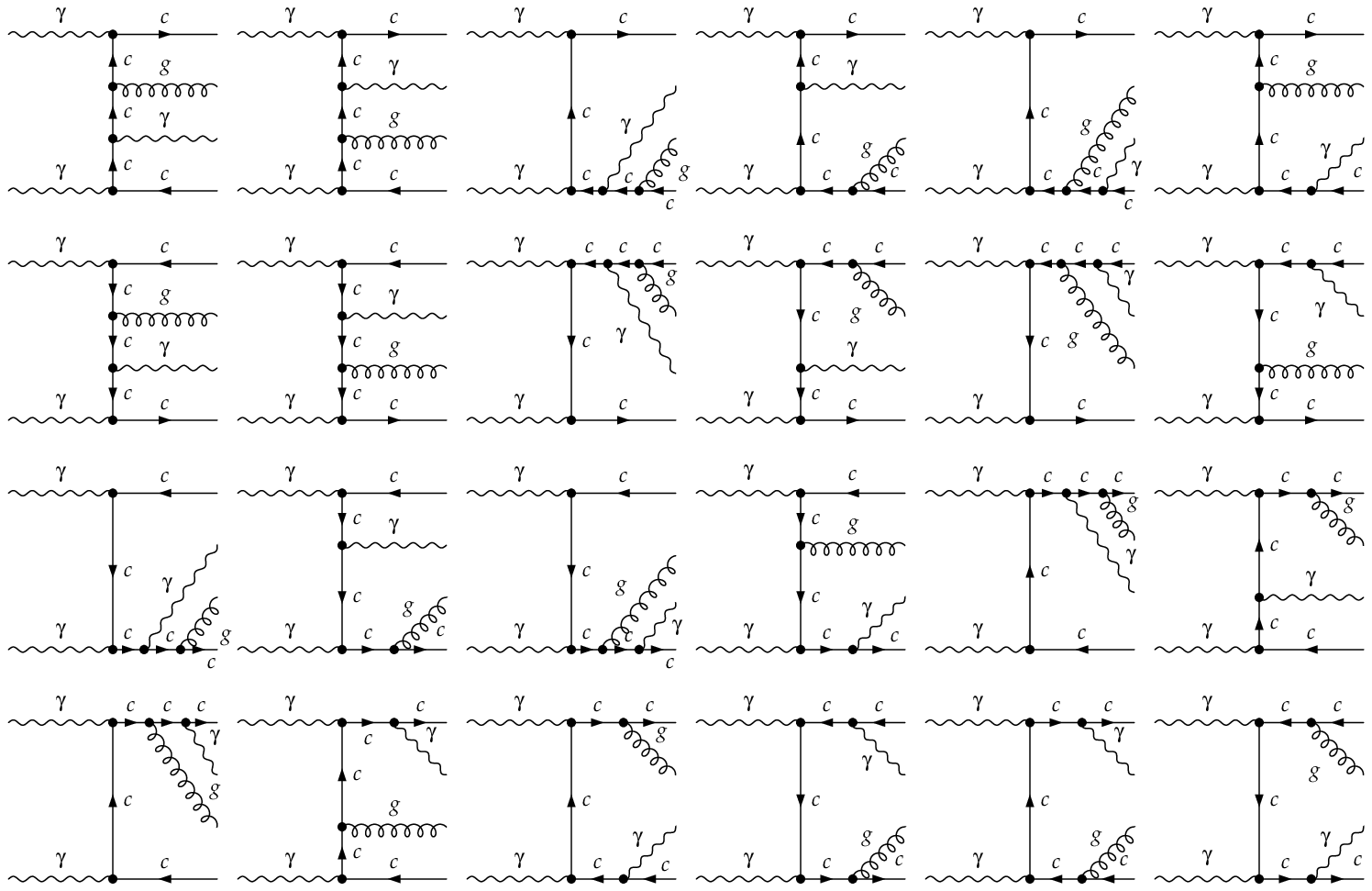}
\caption{Tree-level diagrams pertinent to the partonic
subprocess~(\ref{eq:ccgy}).}
\label{fig:ccgy}
\end{center}
\end{figure}

\newpage
\begin{figure}[ht]
\begin{center}
\epsfig{figure=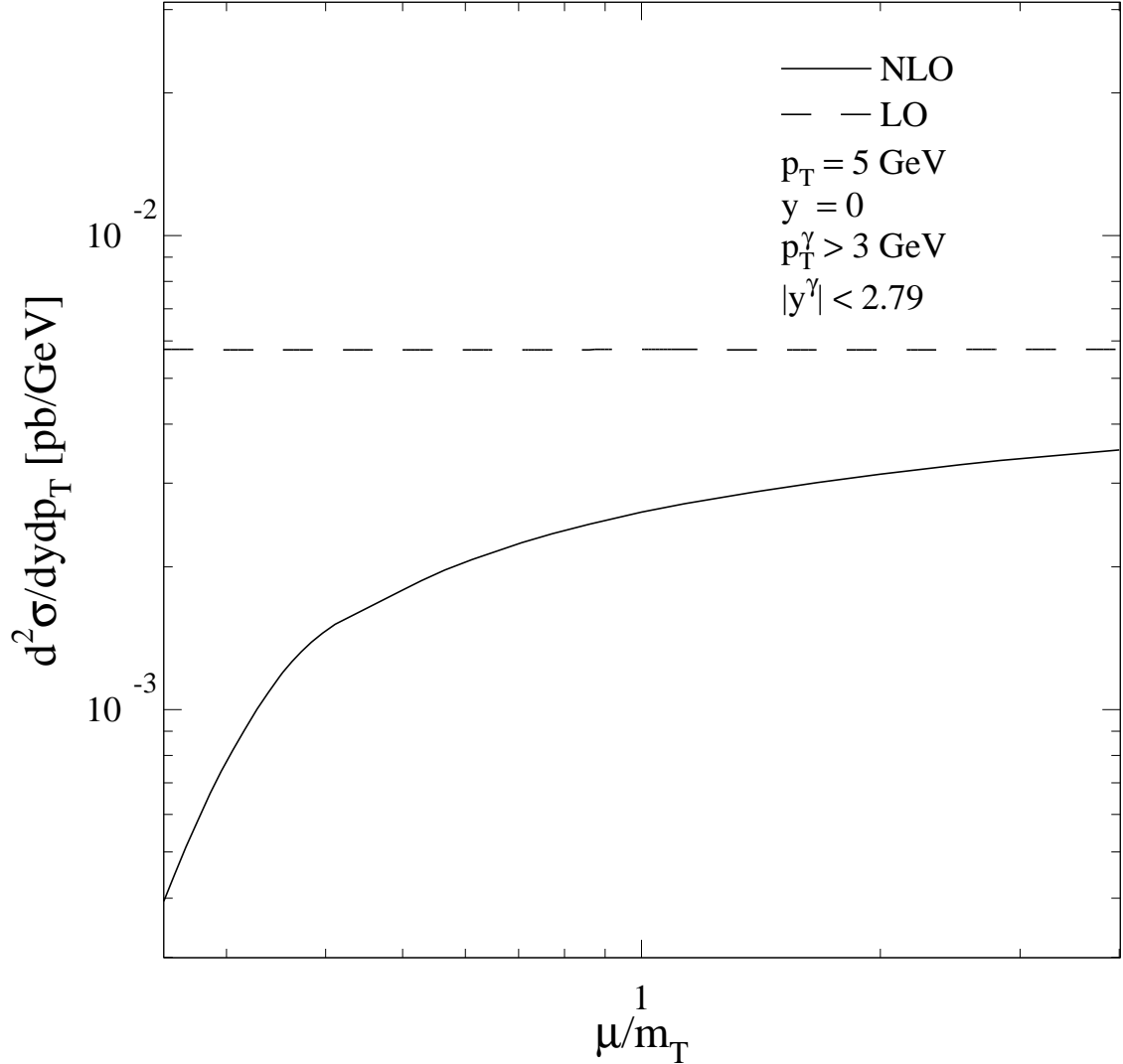,width=\textwidth}
\caption{Differential cross section $d^2\sigma/dp_T\,dy$ in pb/GeV of
$e^+e^-\to e^+e^-J/\psi+X_\gamma$ in direct photoproduction at TESLA with
$\sqrt s=500$~GeV for a prompt $J/\psi$ meson with $p_T=5$~GeV and $y=0$
accompanied by a prompt photon with $p_T^\gamma>3$~GeV and $|y^\gamma|<2.79$.
The LO (dashed line) and NLO (solid line) results are shown as functions of
$\mu$.}
\label{fig:mu}
\end{center}
\end{figure}

\newpage
\begin{figure}[ht]
\begin{center}
\epsfig{figure=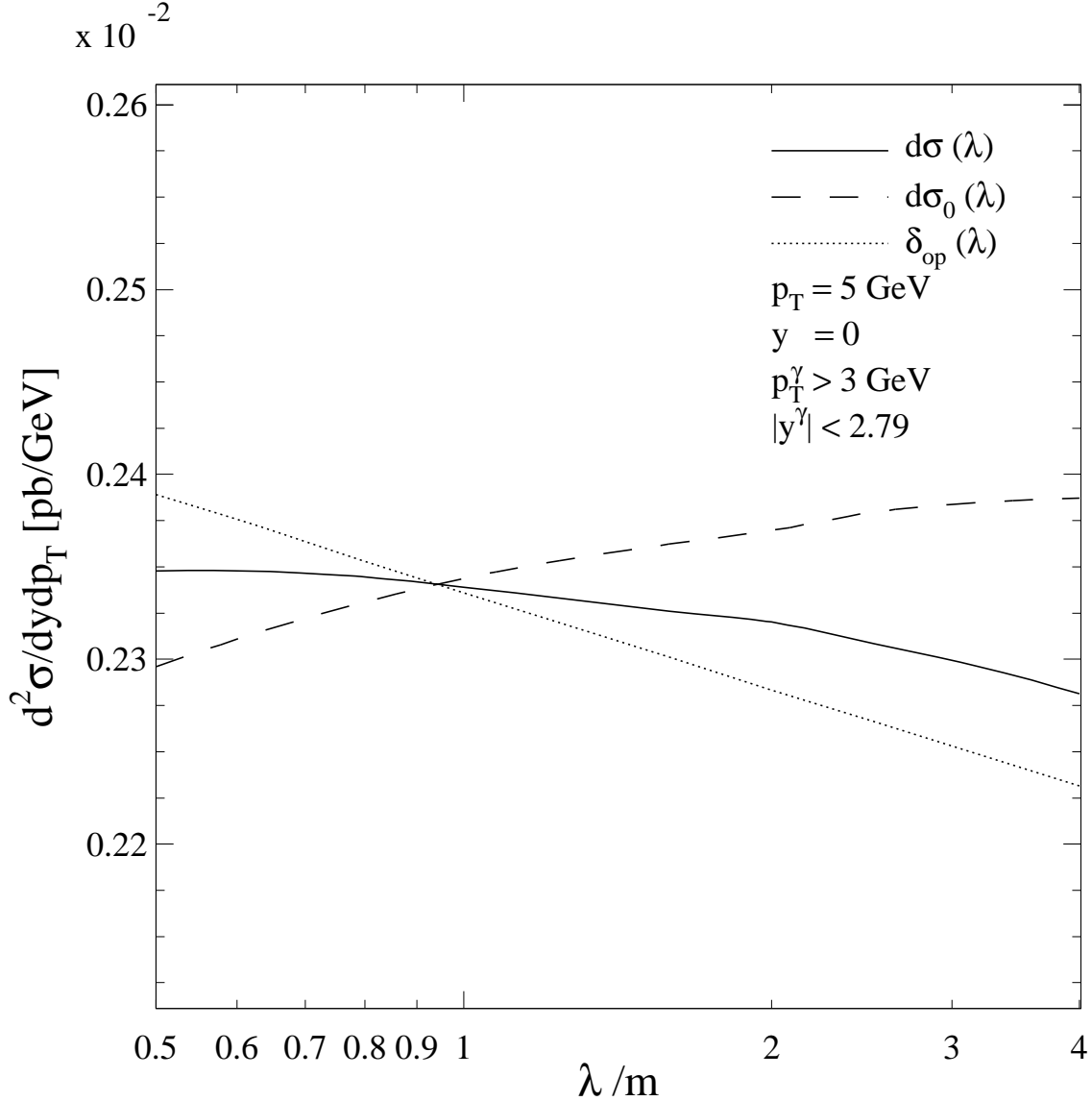,width=\textwidth}
\caption{Differential cross section $d^2\sigma/dp_T\,dy$ in pb/GeV of
$e^+e^-\to e^+e^-J/\psi+X_\gamma$ in direct photoproduction at TESLA with
$\sqrt s=500$~GeV for a prompt $J/\psi$ meson with $p_T=5$~GeV and $y=0$
accompanied by a prompt photon with $p_T^\gamma>3$~GeV and $|y^\gamma|<2.79$.
The NLO result is shown as a function of $\lambda$, (i) keeping $\lambda$ in
the partonic cross sections fixed (dashed line), (ii) keeping $\lambda$ in
$\left\langle{\cal O}^H\left[{}^3\!S_1^{(8)}\right]\right\rangle_r(\lambda)$
fixed (dotted line), and (iii) varying all occurrences of $\lambda$
simultaneously (solid line).}
\label{fig:la}
\end{center}
\end{figure}

\newpage
\begin{figure}[ht]
\begin{center}
\epsfig{figure=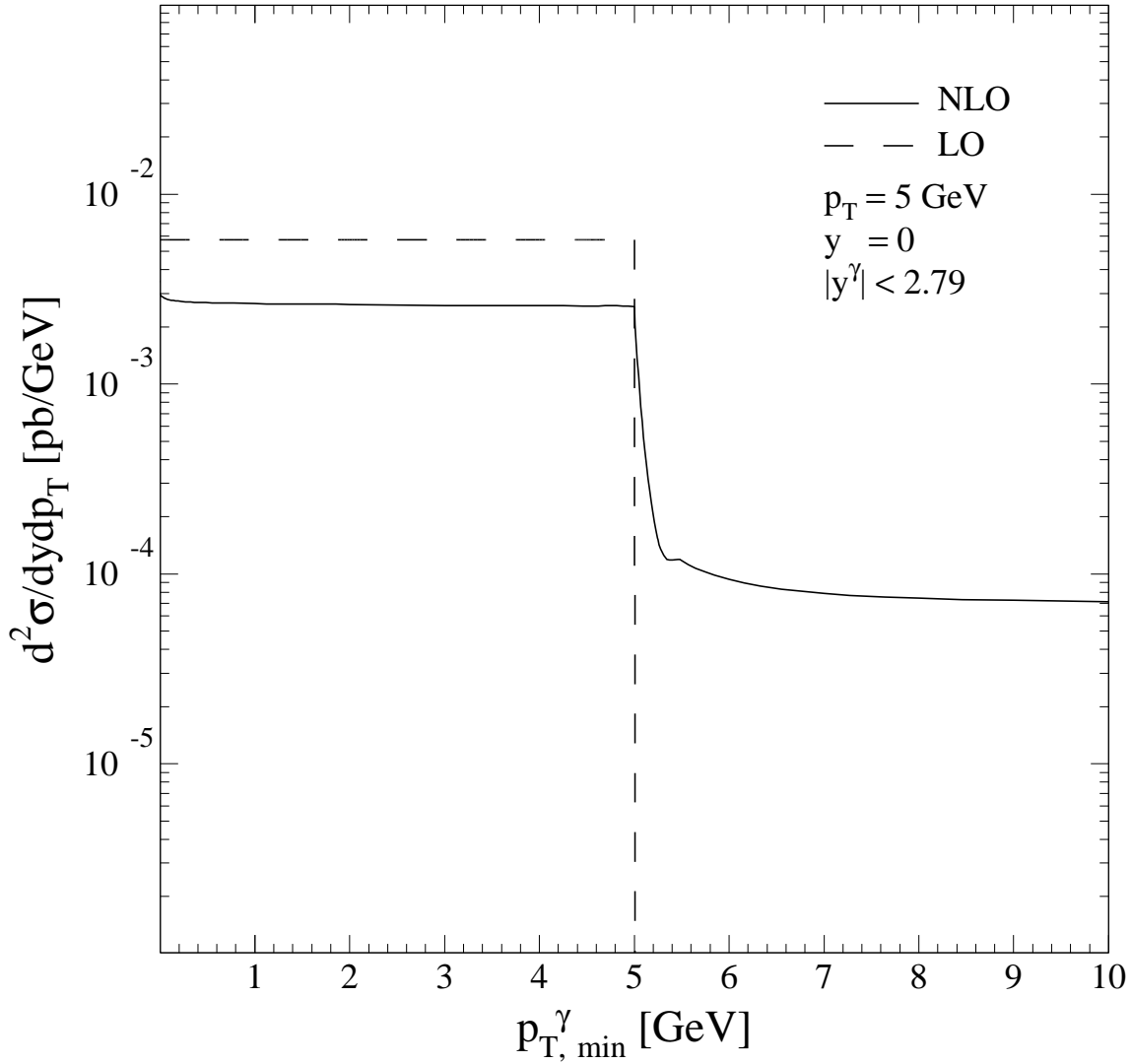,width=\textwidth}
\caption{Differential cross section $d^2\sigma/dp_T\,dy$ in pb/GeV of
$e^+e^-\to e^+e^-J/\psi+X_\gamma$ in direct photoproduction at TESLA with
$\sqrt s=500$~GeV for a prompt $J/\psi$ meson with $p_T=5$~GeV and $y=0$
accompanied by a prompt photon with $p_T^\gamma>p_{T,\,{\rm min}}^\gamma$ and
$|y^\gamma|<2.79$ as a function of $p_{T,\,{\rm min}}^\gamma$ at LO (dashed
line) and NLO (solid line).}
\label{fig:pty}
\end{center}
\end{figure}

\newpage
\begin{figure}[ht]
\begin{center}
\epsfig{figure=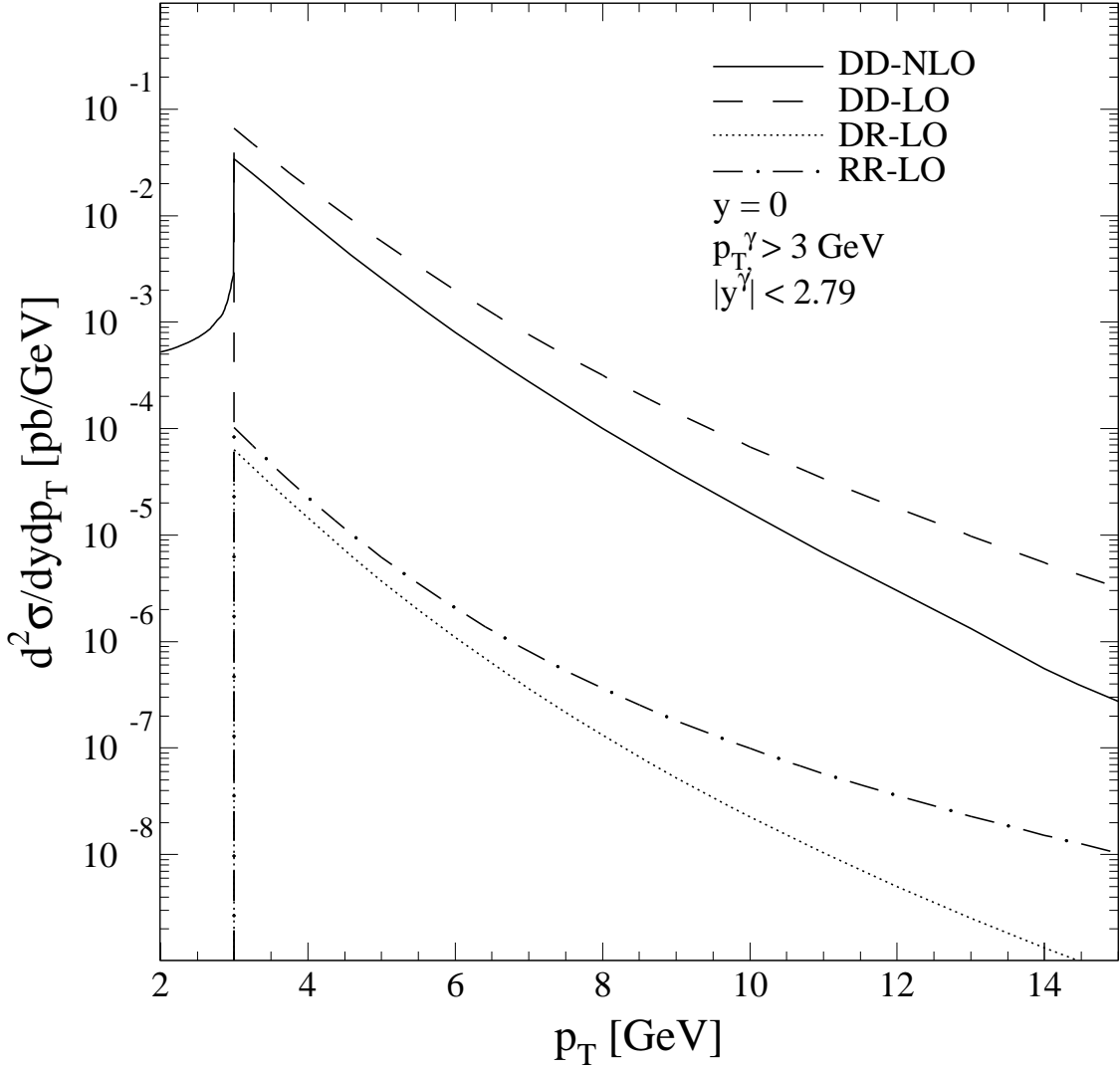,width=\textwidth}
(a)
\caption{Differential cross section $d^2\sigma/dp_T\,dy$ in pb/GeV of
$e^+e^-\to e^+e^-J/\psi+X_\gamma$ at TESLA with $\sqrt s=500$~GeV for a prompt
$J/\psi$ meson accompanied by a prompt photon with $p_T^\gamma>3$~GeV and
$|y^\gamma|<2.79$ (a) for $y=0$ as a function of $p_T$ and (b) for $p_T=5$~GeV
as a function $y$ at LO (dashed line) and NLO (solid line).
In part (a), the LO results of single-resolved (dotted line) and
double-resolved (dot-dashed line) photoproduction are also shown.}
\label{fig:xs}
\end{center}
\end{figure}

\newpage
\begin{figure}[ht]
\begin{center}
\epsfig{figure=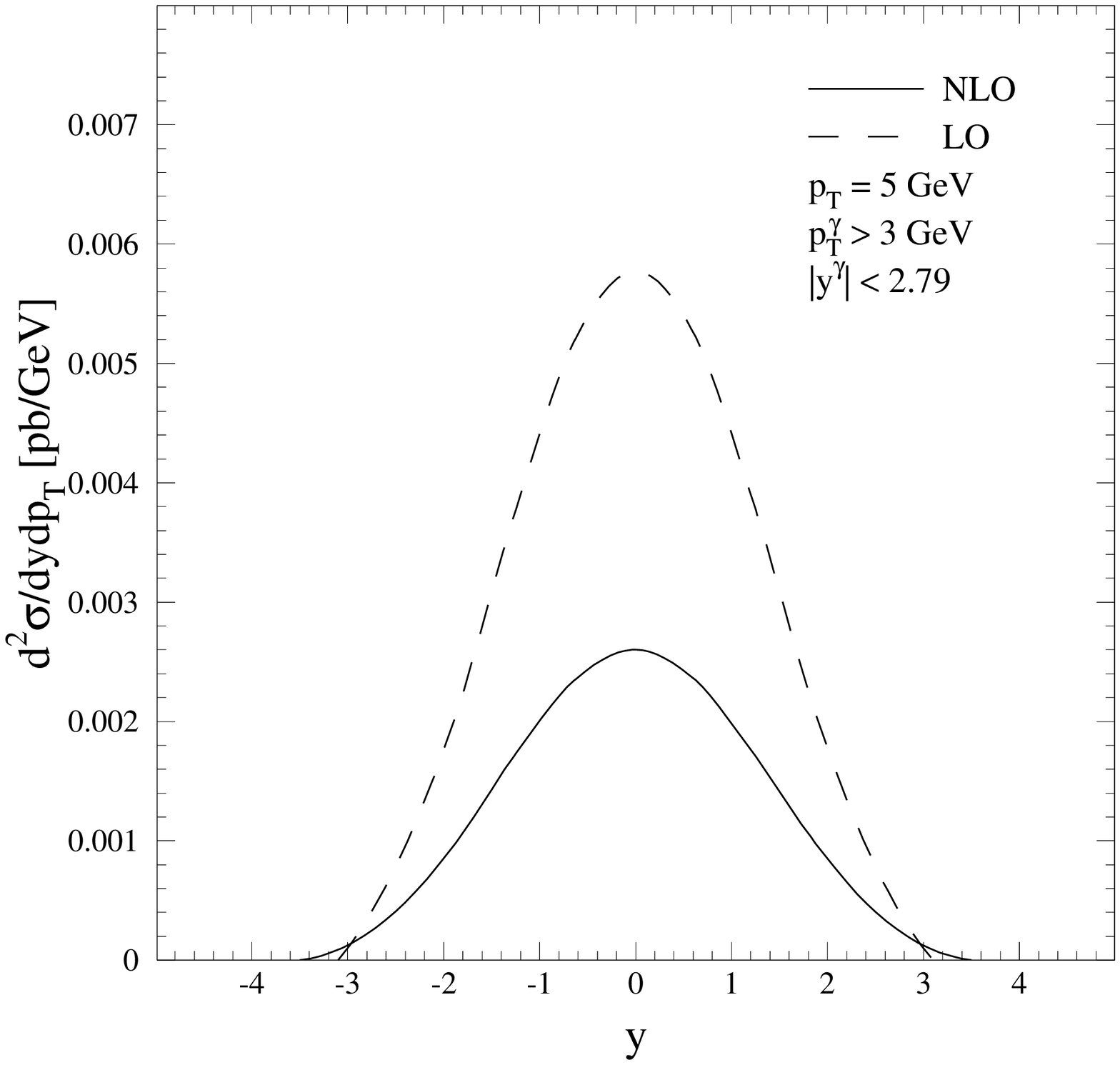,width=\textwidth}
(b)\\
Fig.~\ref{fig:xs} (continued).
\end{center}
\end{figure}

\newpage
\begin{figure}[ht]
\begin{center}
\epsfig{figure=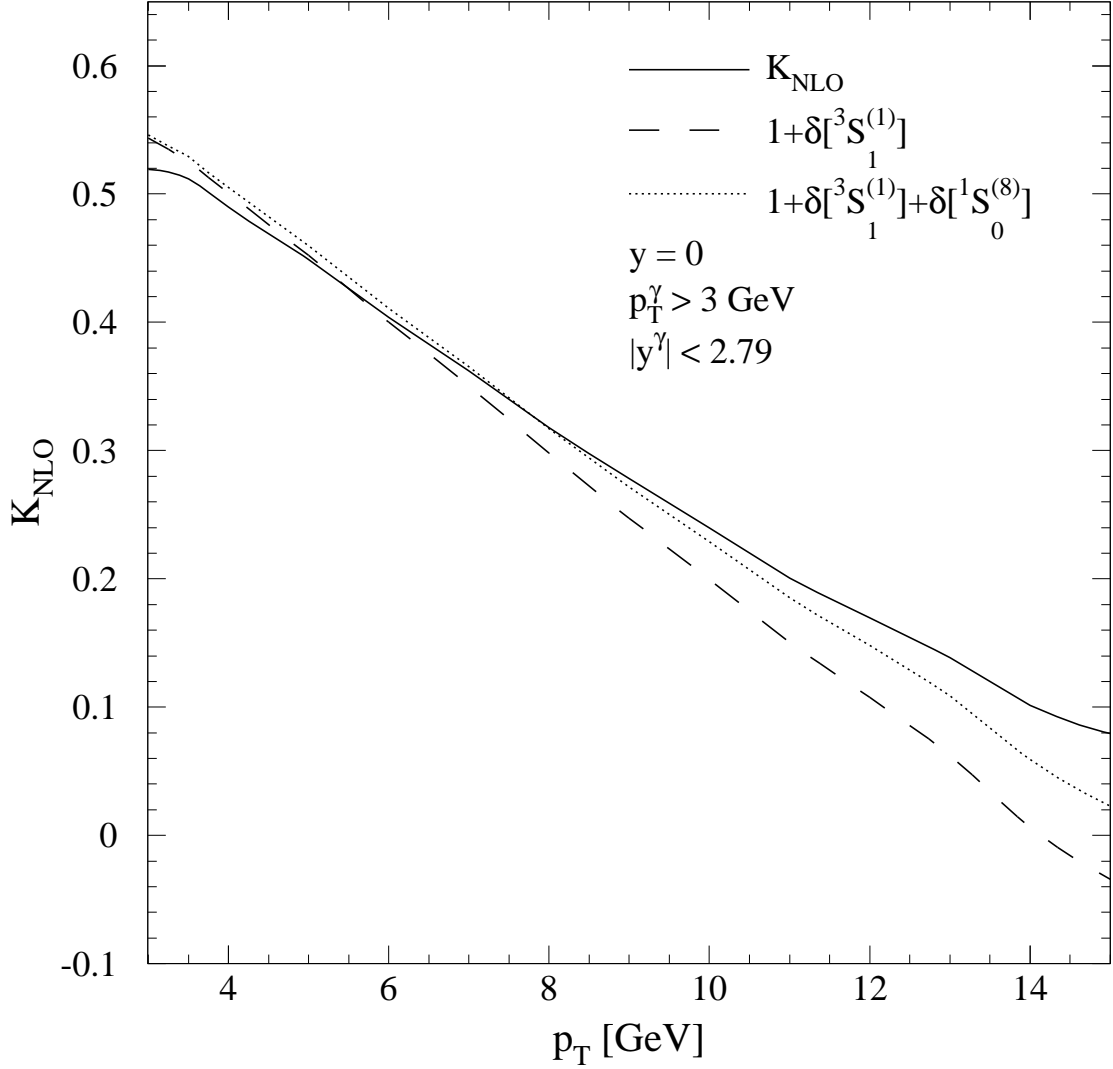,width=\textwidth}
(a)
\caption{QCD correction factor $K$ of the differential cross section
$d^2\sigma/dp_T\,dy$ in pb/GeV of $e^+e^-\to e^+e^-J/\psi+X_\gamma$ at TESLA
with $\sqrt s=500$~GeV for a prompt $J/\psi$ meson accompanied by a prompt
photon with $p_T^\gamma>3$~GeV and $|y^\gamma|<2.79$ (a) for $y=0$ as a
function of $p_T$ and (b) for $p_T=5$~GeV as a function $y$.
The incremental shifts $1+\delta\left[{}^3\!S_1^{(1)}\right]$ (dashed lines)
and $1+\delta\left[{}^3\!S_1^{(1)}\right]+\delta\left[{}^1\!S_0^{(8)}\right]$
(dotted lines) are shown separately.}
\label{fig:k}
\end{center}
\end{figure}

\newpage
\begin{figure}[ht]
\begin{center}
\epsfig{figure=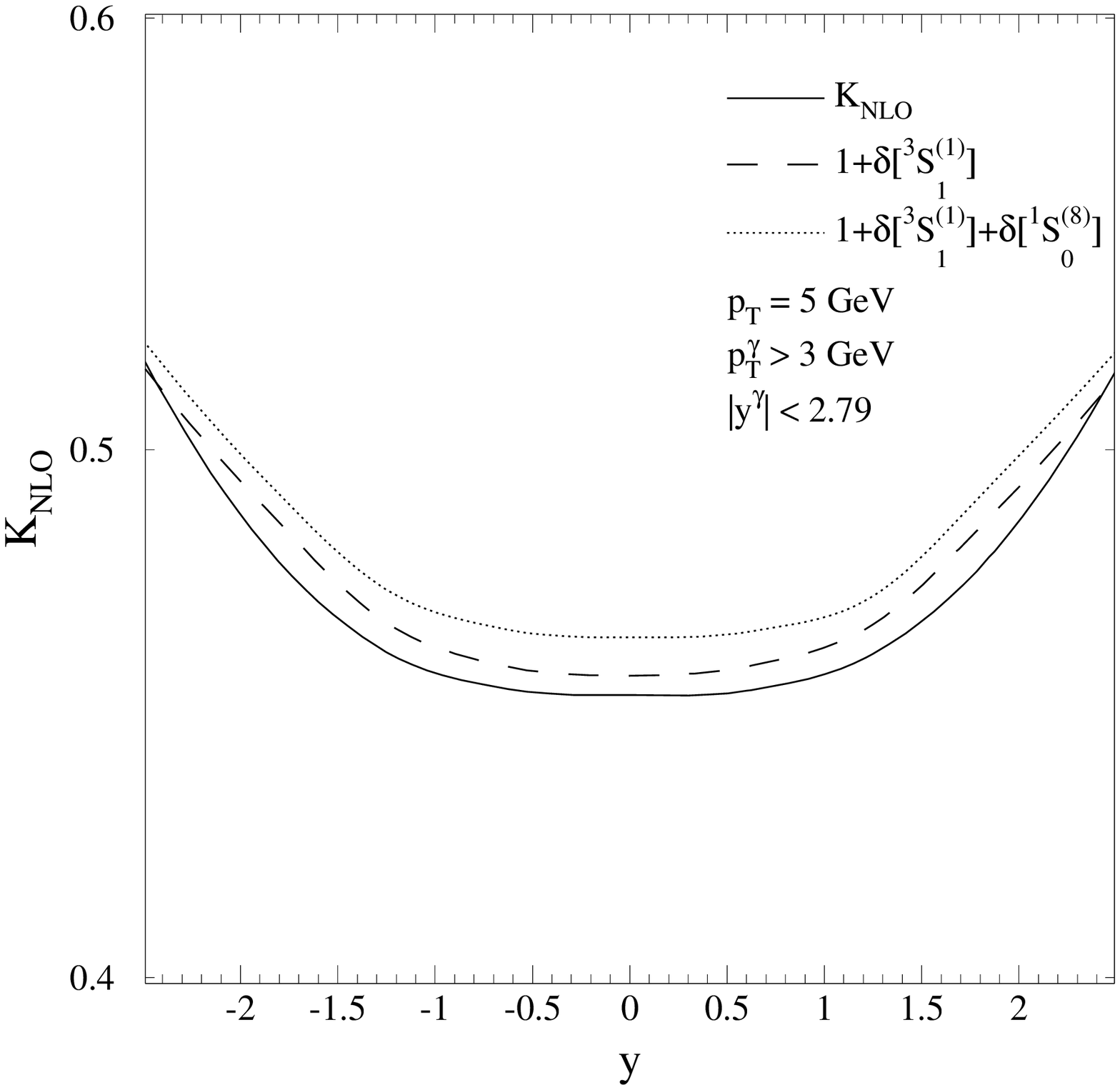,width=\textwidth}
(b)\\
Fig.~\ref{fig:k} (continued).
\end{center}
\end{figure}

\end{document}